\documentclass[aps,pra,groupedaddress,floats,reprint]{revtex4-2}
\usepackage[english]{babel}

\usepackage{amsfonts,graphicx,soul,natbib,mathrsfs}

\usepackage[colorlinks=true]{hyperref}

\usepackage{bm}
\usepackage{epstopdf}
\usepackage{textcomp}
\usepackage{xcolor}
\usepackage{soul}
\usepackage{physics}
\usepackage{graphics}
\usepackage{comment}
\usepackage{amssymb}

\usepackage{dcolumn}
\usepackage{bm}

\begin{document}

\title{Theoretical consideration of a twisted atom}

\author{P. K. Maslennikov}%
\affiliation{School of Physics and Engineering,
ITMO University, 197101 St. Petersburg, Russia}%

\author{A. V. Volotka}%
\affiliation{School of Physics and Engineering,
ITMO University, 197101 St. Petersburg, Russia}%

\author{S. S. Baturin}%
\affiliation{School of Physics and Engineering,
ITMO University, 197101 St. Petersburg, Russia}%

\date{\today}

\begin{abstract}
We investigate the twisted state of an atom and the possible effect of such a state on the properties of the photons emitted as a result of an electron transition in that atom. We first propose a framework for describing the twisted atomic state, and then explore possible differences in the nuclear recoil effects in the twisted atom compared to those in the plane-wave atom. We conclude that if the initial atomic state is twisted, then the photon distribution is altered. We point out that in a certain observation scheme, one can detect a feature of this twist in the distribution of the emitted photons, even in zero order in $m/M$.   
\end{abstract}

\maketitle

\vfill
\section{Introduction}


Structured light, photons with the phase vortex or twisted photons, is a wide and well developed field of study \cite{Allen1,Allen1999, FrankeArnold2008, ct3, TL2, TW3, TL4, UFN}. The concept of a vortex phase has been extended by the duality principle to electrons \cite{Bliokh2007, ushida2010, McMorran2011, Bliokh2012}, neutrons \cite{ct8, Sarenac_2018, jach2022} as well as to a composite quantum system such as atoms and molecules \cite{Lembessis2014}. In the context of atomic physics, investigations have focused mainly on the interaction of twisted light with ``standard'' atoms. In particular, it has already been shown a clear difference in photo-ionization and scattering processes \cite{Hayrapetyan_2013,Har2,Matula_2013, surzhykov2015, surzhykov:2016:033420, karlovets2017, LightandAtom, peshkov2018, afanasev2018, zaytsev2018, forbes2019, afanasev2020, schulz2020, ivanov2022, serbo2022}. In turn, it has been shown experimentally that twisted photons can excite forbidden transitions when selection rules for the electron transitions in the photo-ionization process are modified \cite{schmiegelow2016, ForbidTrans1}. In addition, some theoretical and experimental studies have pointed to the possibility of orbital angular momentum (OAM) transfer from photons to atoms in the photon absorption process \cite{Davis_2013, stopp2022, peshkov2023}.

Recently, it has been experimentally demonstrated \cite{VortAt2021} how to create an entire atom in the vortex state.
In this experiment, the beam of helium atoms was passed through a fork diffraction grating. As a result, the diffracted atoms formed the ring intensity profile, one of the hallmarks of the nonzero OAM quantum state. In view of this experimental progress and considerable theoretical interest in the subject, in this paper, we study the twisted atom and the possible effect of the twist of the atomic state on the properties of the emitted photons in electron transitions. We consider a twisted atom as a twist of the center of mass and explore the interaction between this twist and the electron subsystem through nuclear recoil.

 In our present study, we consider the photon emission process in a hydrogen-like atom. We study how the initial twisted state of the center of mass of the atom affects the $S$ matrix, the transition probabilities, and the photon distribution. We computed the $S$ matrix of the single-photon emission due to the electron transition for three different cases: when both the initial and final states of the center of mass are plane waves, when both states are twisted, and when the initial state is twisted and the final state is a plane wave. We show that in a common scenario where the final state of the atom is not detected and the transverse momenta of the center of mass is small, the reduced differential probability is somewhat similar to the commonly known result. However, if the latter is not the case we show that the differential probability of the photon emission is different and potentially can be experimentally detected if the opening angle $\theta=\arctan(P_{\perp}/P_z)$ of the twisted center of mass state is reasonably large. 
On top of that we propose a special experiment with the coincidence scheme detector that can also reveal the initial twisted nature of the center of mass when the final atomic state is projected onto a plane wave and the distribution of the emitted photons is simultaneously measured. However, the information about the orbital angular momentum of the atom is lost in this measurement.

Throughout the paper we use relativistic units ($\hbar = c = 1, e < 0$).


\section{Electron-nucleus Hamiltonian}


We consider the nonrelativistic Hamiltonian of the hydrogen-like atom interacting with the second-quantized electromagnetic radiation field in the transverse gauge. In  the Schr\"odinger representation it can be written as \cite{bethe}
\begin{eqnarray}
\label{eq:mainH}
\hat{\mathcal{H}} &=& \frac{\left[\hat{\mathbf{p}}_e-e\hat{\mathbf{A}}(t,{\bf r}_e)\right]^2}{2m} + \frac{\left[\hat{\mathbf{p}}_n + eZ\hat{\mathbf{A}}(t,{\bf r}_n)\right]^2}{2M}\nonumber\\
&+& V(|\mathbf{r}_n-\mathbf{r}_e|)+W_f,
\end{eqnarray}
where $W_f$ is the external field energy, $V$ is the electron-nucleus interaction potential, $\hat{\mathbf{A}}(t, {\bf r})$ is the (transverse) vector potential of the quantized electric ($\hat{\mathbf{E}} = -\partial_t \hat{\mathbf{A}}$) and magnetic ($\hat{\mathbf{H}} = \nabla\times\hat{\mathbf{A}}$) fields. The following notation is introduced above: $m$ - electron mass, $M$ - mass of the nucleus, $Z$ - charge number of the nucleus, index $e$ stands for the electron momentum and coordinate, and index $n$ stands for the nucleus momentum and coordinate.
We note that this particular Hamiltonian is the nonrelativistic limit of the Breit equation with omitted spin interactions and orbital coupling \cite{BLP}.

To identify the coordinates of an atom as a whole, we switch to the coordinates of the center of mass (see for example  Ref.\cite{Shabaev2001}):
\begin{align}
\label{eq:coord}
&\mathbf{R} = \frac{\mathbf{r}_em+\mathbf{r}_nM}{m+M}, \nonumber\\
&\mathbf{r} = \mathbf{r}_e - \mathbf{r}_n.
\end{align}
Momentum transforms as follows  
\begin{align}
\label{eq:mom}
    &\hat{\mathbf{p}}=\hat{\mathbf{p}}_e-\frac{m}{m+M}\left(\hat{\mathbf{p}}_e+\hat{\mathbf{p}}_n \right), \nonumber \\
    &\hat{\mathbf{P}}=\hat{\mathbf{p}}_e+\hat{\mathbf{p}}_n.
\end{align}
Substituting Eq.~\eqref{eq:coord} and Eq.~\eqref{eq:mom} into the Eq.~\eqref{eq:mainH} and decomposing in series assuming $m/M \ll 1$ we get in the zero order in $m/M$ 
\begin{align}
\label{eq:HHz}
\hat{\mathcal{H}} = \hat{\mathcal{H}}_0 + \hat{\mathcal{H}}_i + \mathcal{O}\left[\frac{m}{M} \right].
\end{align}
The unperturbed Hamiltonian $\hat{\mathcal{H}}_0$ reads
\begin{align}
\label{eq:H0}
    \hat{\mathcal{H}}_0 = \frac{\hat{\mathbf{p}}^2}{2m}+\frac{\hat{\mathbf{P}}^2}{2M}+V(r)+W_f
\end{align}
and the interaction Hamiltonian has the form
\begin{eqnarray}
\label{eq:Hint}
\hat{\mathcal{H}}_i &=& -\frac{e}{m}\mathbf{\hat{p}}\hat{\mathbf{A}}(t,{\bf R}+{\bf r}) - \frac{e}{M}\mathbf{\hat{P}}\hat{\mathbf{A}}(t,{\bf R+r})\nonumber\\
&+&\frac{eZ}{M}\hat{\mathbf{P}}\hat{\mathbf{A}}(t,\mathbf{R}).
\end{eqnarray}
Above, we keep only the terms linear in $\hat{\mathbf{A}}$, since we are going to consider single photon process only. Inclusion of higher orders requires inclusion of the relativistic corrections as well. In the present study we focus on the most simple case that already shows some difference between the plane wave and twisted wave states.
We note that the inclusion of the spin and consideration of a multielectron atom do not affect the further analysis, so we omit common terms such as electron-electron interaction and spin for simplicity. Moreover, we restrict ourselves to the zero order in $m/M$, while the higher-order corrections can be accounted for by perturbation theory; see e.g. for the transition amplitude, Refs.~\cite{fried:1963:574, bacher:1984:135, karshenboim:1997:4311, shabaev:1998:907, pachucki:2003:012504, volotka:2008:167, bondy:2020:052807}. 

We stress that the Hamiltonian \eqref{eq:HHz} is limited to the zero order in $m/M$ only and all further analysis do not include higher order effects. Interestingly, if the interaction with the electromagnetic fields is limited to the dipole approximation one may benefit from the Hamiltonian derived in Ref.\cite{pachucki:2008} that is valid in all orders in $m/M$. The analysis of the latter should not differed in principle from the analysis of the Hamiltonian Eq.\eqref{eq:HHz} with the interaction given by \eqref{eq:H0}.

As one can see from Eq.~\eqref{eq:H0}, the center of mass and the relative electron variables are separated, and therefore the full wave function is represented as the product of the wave function of the electron subsystem, $|\phi \rangle$, and the wave function of the center of mass, $|\Phi \rangle$, as
\begin{equation}
\label{eq:wavefactor}
    |\Phi, \phi \rangle = |\Phi \rangle  |\phi \rangle,
\end{equation}
with
\begin{equation}
\label{eq:com}
\frac{\hat{\mathbf{P}}^2}{2M}|\Phi\rangle = E |\Phi\rangle,
\end{equation}
and
\begin{equation}
\left[\frac{\hat{\mathbf{p}}^2}{2m}+V(r)\right] |\phi \rangle = \varepsilon |\phi\rangle
\end{equation}
where $E$ and $\varepsilon$ are the center of mass and electron energies. Eq.~\eqref{eq:com} describes the motion of an atom as a whole and, thus, its solution characterizes the properties of the beam. In what follows we consider two cases: plane and twisted beams of atoms.


\section{Single photon process}

\begin{figure}[t]
    \centering
     \includegraphics[width=0.5\textwidth]{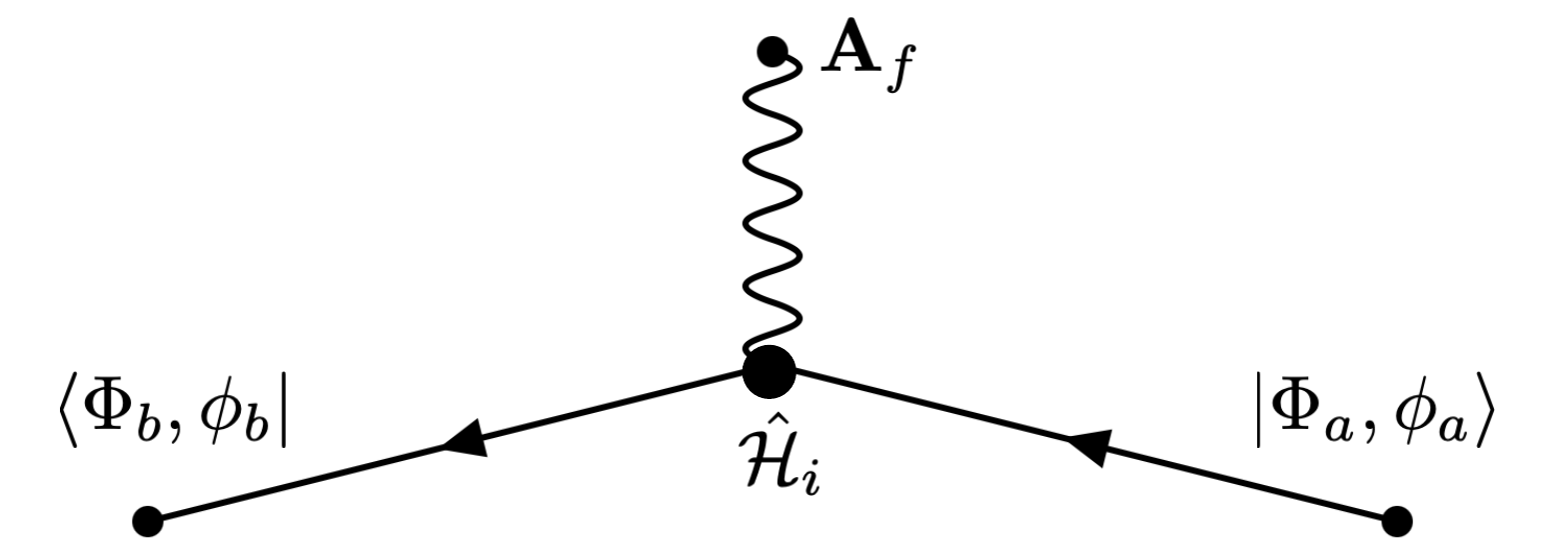}
    \caption{Feynman diagram corresponding to the lowest order interaction of the atom with the quantized electromagnetic field. The atom emits a photon as a result of the electron transition and experiences recoil.}
    \label{fig:diag}
\end{figure}

The $S$-matrix of the transition of the atom from state $a$ to state $b$ with the emission of the photon ($f$) with the wave vector $\mathbf{k}_p$, energy $\omega=|\mathbf{k}_p|$ and polarization vector $\boldsymbol{\epsilon}_p$ is given by the following scalar product (see Fig.~\ref{fig:diag}) \cite{BLP}
\begin{eqnarray}
\label{eq:Smgen0}
S = -i\int_{-\infty}^{\infty} dt \langle f | \langle \Phi_b,\phi_b |\hat{\mathcal{H}}_i|\Phi_a,\phi_a\rangle |0\rangle,
\end{eqnarray}
where $|f\rangle$ and $|0\rangle$ are the photon Fock states with one photon (with all quantum numbers notated as $f$) and zero photons. Expanding the photon field operator in terms of annihilation (creation) operators $\hat{a}_{f'}$ ($\hat{a}^+_{f'}$) as
\begin{eqnarray}
    \hat{\mathbf{A}}(t,{\bf r}) = \sum_{f'} \left[\hat{a}_{f'} \mathbf{A}_{f'}(t,{\bf r}) + \hat{a}^+_{f'} \mathbf{A}^*_{f'}(t,{\bf r})\right]
\end{eqnarray}
where $\mathbf{A}_f$ is the photon wave function, and substituting it into Eq.~\eqref{eq:Smgen0} one gets
\begin{eqnarray}
\label{eq:Smgen}
S &=& -i\int_{-\infty}^{\infty} dt e^{it(\varepsilon_a+E_a-\varepsilon_b-E_b-\omega)} \nonumber\\
&\times& \langle \Phi_b,\phi_b |\hat{\mathcal{H}}_i^f|\Phi_a,\phi_a\rangle,
\end{eqnarray}
here $\hat{\mathcal{H}}_i^f$ is the Hamiltonian $\hat{\mathcal{H}}_i$ where the second-quantized electromagnetic radiation field $\hat{\mathbf{A}}(t,{\bf r})$ is replaced by the coordinate part of the photon wave function $\mathbf{A}_f^*({\bf r})$.

Further, with the help of the identity
\begin{align}
    \int_{-\infty}^\infty dt &e^{it(\varepsilon_a+E_a-\varepsilon_b-E_b-\omega)}=\nonumber \\&2\pi \delta(\varepsilon_a+E_a-\varepsilon_b-E_b-\omega),
\end{align}
we have
\begin{eqnarray}
\label{eq:S}
S &=& -2\pi i \delta(\varepsilon_a+E_a-\varepsilon_b-E_b-\omega)\nonumber\\
&\times&\langle \Phi_b,\phi_b,f |\hat{\mathcal{H}}_i^f|\Phi_a,\phi_a\rangle.
\end{eqnarray}
Here capital letters ($\Phi_a$, $\Phi_b$, $E_a$, $E_b$) correspond to the state of the center of mass and small letters ($\phi_a$, $\phi_b$, $\varepsilon_a$, $\varepsilon_b$) correspond to the state of the electron subsystem.
We consider two differential amplitudes $dw$ and $dw_r$. The first case of $dw$ is when all initial and final states are detected: the state of the emitted photon, the initial and final states of the electron subsystem, and the initial and final states of the center of mass. In this case, the differential probability per unit time is
\begin{equation}
    dw = \frac{|S|^2}{T}dn_b dn_p,
\end{equation}
where $dn_b$ and $dn_p$ are the number of states for the center of mass and emitted photon in the given phase-space volumes. Regularizing the square of the energy $\delta$-function in a common way
\begin{align}
    \delta(\varepsilon_a+E_a-&\varepsilon_b-E_b-\omega)^2=\\ \nonumber &\frac{T}{2\pi}\delta(\varepsilon_a+E_a-\varepsilon_b-E_b-\omega)
\end{align}
we get to
\begin{eqnarray}
\label{eq:prob}
dw&=&2 \pi \delta(\varepsilon_a+E_a-\varepsilon_b-E_b-\omega) 
\nonumber\\
&\times& |\langle\Phi_b,\phi_b,f|\hat{\mathcal{H}}_i^f|\Phi_a,\phi_a\rangle|^2 dn_b dn_p.
\end{eqnarray}
Above we have taken into account that the density of states for the bound electron is unity. The second case of $dw_r$, which is the most common in spectroscopic measurements, occurs when the final state of the atom is not measured and, therefore, the final state of the center of mass is not detected. The reduced probability for the latter could be found by integrating over Eq.~\eqref{eq:prob} the final state of the center of mass
\begin{eqnarray}
\label{eq:probR}
dw_r &=& 2\pi dn_p \int dn_b\,\delta(\varepsilon_a+E_a-\varepsilon_b-E_b-\omega) \nonumber\\
&\times&|\langle \Phi_b,\phi_b,f|\hat{\mathcal{H}}_i^f|\Phi_a,\phi_a\rangle|^2.   
\end{eqnarray}


\section{Plane-wave basis}


First we reproduce the case when the center of mass is described by a plane wave (see Refs.~\cite{BLP, Peskin}) and the interaction Hamiltonian is given by Eq.~\eqref{eq:Hint}. In this case the solution of Eq.~\eqref{eq:com} is given by
\begin{align}
\label{eq:PWF}
    |\Phi^{\rm PW} \rangle = \frac{1}{\sqrt{2 E V}}
     \exp \left(i \mathbf{P} \mathbf{R}\right),
\end{align}
here
\begin{align}
    E=\frac{P^2}{2M}
\end{align}
and $V$ is the normalization volume. Plane wave states are normalized to a delta function as
\begin{align}
     \langle \Phi^{\rm PW}_b|\Phi^{\rm PW}_a \rangle =  
     \frac{(2\pi)^3}{2E_a V}\delta^3(\mathbf{P}_a-\mathbf{P}_b).
\end{align}
Moreover, we assume that the emitted photon is described by the plane wave and that the coordinate part of the photon wave function is given by
\begin{align}
\mathbf{A}_{f}(\mathbf{r}) \equiv \mathbf{A}_{{\bf k}\Lambda}(\mathbf{r}) = \frac{1}{\sqrt{2\omega V}}\boldsymbol{\epsilon}_\Lambda\exp (i\mathbf{k}\mathbf{r}).
\end{align}
It is apparent that 
\begin{align}
\label{eq:photfact} \mathbf{A}_f(\mathbf{r}+\mathbf{R}) = \mathbf{A}_f(\mathbf{r})e^{i\mathbf{k}_p\mathbf{R}} = \mathbf{A}_f(\mathbf{R})e^{i\mathbf{k}_p\mathbf{r}}.
\end{align}
Assuming further the dipole approximation (${\bf k}_p {\bf r} \ll 1$) we simplify
\begin{equation}
    \mathbf{A}_f(\mathbf{r}+\mathbf{R}) \simeq \mathbf{A}_f(\mathbf{R})
\end{equation}
and, thus, the second and third terms of Eq.~\eqref{eq:Hint} do not contribute to the amplitude. Consequently, substituting first term of Eq.~\eqref{eq:Hint} into Eq.~\eqref{eq:S} the $S$-matrix can be written as the product of the electron matrix element and the center of mass matrix element as follows
\begin{eqnarray}
    S&=&\frac{2\pi i}{\sqrt{2\omega V}}\frac{e}{m}\delta(\varepsilon_a+E_a-\varepsilon_b-E_b-\omega) \nonumber \\ 
    &\times&\langle \Phi^{\rm PW}_b|e^{-i\mathbf{k}_p\mathbf{R}}| \Phi^{\rm PW}_a \rangle 
    \langle \phi_b| \boldsymbol{\epsilon}_p \hat{\mathbf{p}} |\phi_a\rangle.   
\end{eqnarray}
The center of mass matrix element evaluates to
\begin{align}
    \langle \Phi^{\rm PW}_b|e^{-i\mathbf{k}_p\mathbf{R}}| \Phi^{\rm PW}_a \rangle = \frac{(2\pi)^3}{2\sqrt{E_a E_b} V}\delta^3(\mathbf{P}_a-\mathbf{P}_b-\mathbf{k}_p),
\end{align}
and finally the S-matrix in the case of the plane wave initial and final states of the center of mass and final photon state reads
\begin{eqnarray}
    S^{\rm PW} &=& \frac{(2\pi)^4 i}{2V\sqrt{2\omega E_a E_b V}}\frac{e}{m} \delta(\varepsilon_a+E_a-\varepsilon_b-E_b-\omega) \nonumber\\
    &\times& \delta^3(\mathbf{P}_a-\mathbf{P}_b-\mathbf{k}_p)
     \langle \phi_b| \boldsymbol{\epsilon}_p \hat{\mathbf{p}}|\phi_a\rangle.
\end{eqnarray}
We note that $S^{\rm PW}$ is a product $S^{\rm PW} = S^{\rm PW}_c S_e$ of 
\begin{eqnarray}
\label{eq:Spwc}
    S^{\rm PW}_{c}&=&\frac{(2\pi)^4 i}{2V\sqrt{2\omega E_a E_b V}}
    \delta(\varepsilon_a+E_a-\varepsilon_b-E_b-\omega)\nonumber\\
    &\times&\delta^3(\mathbf{P}_a-\mathbf{P}_b-\mathbf{k}_p)
\end{eqnarray}
that corresponds to the contribution from the integrals over time and center of mass part, and 
\begin{align}
\label{eq:Spwe}
    S_{e}=\frac{e}{m} \langle \phi_b| \boldsymbol{\epsilon}_p \hat{\mathbf{p}}|\phi_a\rangle
\end{align}
that depends only on the initial and the final state of the electron.

In the case of the plane wave final states the number of the states is given by
\begin{eqnarray}
\label{eq:phasePW}
    dn_b &=& \frac{Vd^3P_b}{(2\pi)^3},\nonumber\\
    dn_p &=& \frac{Vd^3k_p}{(2\pi)^3}.
\end{eqnarray}
Substituting the above equations into Eq.~\eqref{eq:prob}, and utilizing the regularization
\begin{align}
\label{eq:Momdl}
    \left[\delta^3(\mathbf{P}_a-\mathbf{P}_b-\mathbf{k}_p) \right]^2 = \frac{V}{(2\pi)^3} \delta^3(\mathbf{P}_a-\mathbf{P}_b-\mathbf{k}_p).
\end{align}
and densities for the final plane-wave states \eqref{eq:phasePW} we get for the differential probability 
\begin{eqnarray}
\label{eq:PPL}
    dw &=& 
      \frac{|S_e|^2}{(2\pi)^2} \frac{1}{2E_a} \delta(\varepsilon_a+E_a-\varepsilon_b-E_b-\omega) \nonumber\\
      &\times&\delta^3(\mathbf{P}_a-\mathbf{P}_b-\mathbf{k}_p)\frac{d^3 k_p}{2\omega} \frac{d^3 P_b}{2E_b}.
\end{eqnarray}
In case when the final state of the center of mass is not detected, we integrate over the final state of the center of mass Eq.~\eqref{eq:probR} and arrive at the reduced probability
\begin{eqnarray}
\label{eq:PredPL}
    dw_r &=&\frac{|S_e|^2}{(2\pi)^2} \frac{1}{2E_a}\delta\left(\varepsilon_a-\varepsilon_b-\omega+\frac{\mathbf{P}_a\mathbf{k}_p}{M}-\frac{k_p^2}{2M} \right) \nonumber \\
    &\times&\left(\frac{P_a^2}{2M}-\frac{\mathbf{P}_a\mathbf{k}_p}{M}+\frac{k_p^2}{2M}\right)^{-1}\frac{d^3 k_p}{4\omega}.
\end{eqnarray}
Without loss of generality one may choose the $z$ -axis along ${\bf P}_a$, such that $P_{z,a} = |{\bf P}_a|$. In this case the Eq.~\eqref{eq:PredPL} takes a form
\begin{eqnarray}
\label{eq:PredPLZ}
    dw_r &=&\frac{|S_e|^2}{(2\pi)^2} \frac{1}{8E_a}\delta\left(\varepsilon_a-\varepsilon_b-\omega+\frac{P_{z,a}\,\omega\cos{\theta_p}}{M}-\frac{\omega^2}{2M} \right) \nonumber \\
    &\times&\left(\frac{P_{z,a}^2}{2M}-\frac{P_{z,a}\omega\cos{\theta_p}}{M}+\frac{\omega^2}{2M}\right)^{-1}\omega d\omega  d \Omega_p.
\end{eqnarray}
where $\Omega_p$ is the solid angle
of the photon emitted during the process considered. The obtained equations \eqref{eq:PPL}, \eqref{eq:PredPL}, and \eqref{eq:PredPLZ} for differential probabilities coincide with known results.


\section{Twisted-wave basis}


We now consider the case of the twisted state of the center of mass in both the initial and final states for the solution of Eq.~\eqref{eq:com}. The twisted wave function is proportional to the Bessel function of the first kind and is given by \cite{Jentschura:2011aa, Ivanov2011, surzhykov:2016:033420, UFN}.
\begin{align}
\label{eq:TWW}
    |\Phi^{\mathrm{TW}} \rangle = \sqrt{\frac{\pi}{RL_z}}\sqrt{\frac{\kappa}{4\pi E}}J_m(\kappa \rho)e^{im\phi+iP_z z}.
\end{align}
where $\kappa$ and $P_z$ are the transverse and longitudinal momenta, $m$ is the projection of the total angular momentum and $E$ is the energy, $E = (\kappa^2+P_z^2)/(2M)$. The wave function given by Eq.~\eqref{eq:TWW} is defined in such a way that in a large but finite cylindrical volume $\pi R^2 L_z$ there is a state of one particle. The twisted wave functions are normalized as follows
\begin{eqnarray}
\langle \Phi^{\mathrm{TW}}_b |\Phi^{\mathrm{TW}}_a \rangle &=& \frac{\pi}{RL_z} \frac{1}{2E_a}\delta(P_{z,a}-P_{z,b})\nonumber\\
&\times&\delta(\kappa_a-\kappa_b)\delta_{m_a m_b}.
\end{eqnarray}

It is convenient to represent a Bessel state as a coherent superposition of plane waves as
\begin{align}
\label{eq:TWW2}
    |\Phi^{\mathrm{TW}} \rangle = \sqrt{\frac{\pi}{RL_z}}\sqrt{\frac{\kappa}{2E}}\frac{1}{(2\pi)^{3/2}}\int\limits_{0}^{2\pi} i^{-m} e^{im\phi} e^{i\mathbf{P}\mathbf{R}}d\phi.
\end{align}
Hence, the center of mass matrix element equals double angular integral of the plane wave matrix element with a phase factor
\begin{align}
\label{eq:melCTW}
     &\langle\Phi_b^{\mathrm{TW}}|e^{-i\mathbf{k}_p\mathbf{R}}|\Phi_a^{\mathrm{TW}}\rangle=  
     \frac{\pi}{RL_z}\frac{\sqrt{\kappa_a\kappa_b}}{2\sqrt{E_a E_b}}
     i^{m_b-m_a}\nonumber\\
     &\times\iint \delta^3(\mathbf{P}_a-\mathbf{k}_p-\mathbf{P}_b)e^{i(m_a\phi_a-m_b\phi_b)} d\phi_ad\phi_b,
\end{align}
here we assume that the photon is described by the plane wave state and only the first term of interaction Hamiltonian \eqref{eq:Hint} contributes. Taking representation of the $\delta$-function in the cylindrical coordinates 
\begin{align}
    \delta^3(\pmb{a}-\pmb{b})=\frac{\delta(|a_\perp|-|b_\perp|)}{|a_\perp|}\delta(a_z-b_z)\delta(\phi_a-\phi_b),
\end{align}
and evaluating one angular integral we get 
\begin{align}
      &\langle\Phi_b^{\mathrm{TW}}|e^{-i\mathbf{k}_p\mathbf{R}}|\Phi_a^{\mathrm{TW}}\rangle = \frac{\pi \delta(P_{z,a}-P_{z,b}-k_{z,p})}{2RL_z\sqrt{E_a E_b}} 
      i^{m_b-m_a} \nonumber \\
      &\times\int \sqrt{\frac{\kappa_b}{\kappa_a}}\delta(\kappa_a-x)e^{i(m_a\phi_x-m_b\phi_b)}d\phi_b. 
      \label{eq:TWM0}
\end{align}
Where the following notations were introduced
\begin{align}
\label{eq:not}
    &\mathbf{P}_b = \mathbf{P}_{z,b}+\pmb{\kappa}_b,~~~
    \mathbf{P}_a = \mathbf{P}_{z,a}+\pmb{\kappa}_a, \nonumber \\
    &\mathbf{k}_p = \mathbf{k}_{z,p}+\pmb{\kappa}_p, \nonumber \\
    &\kappa_b = |\pmb{\kappa}_b|,\,\,\kappa_a = |\pmb{\kappa}_a|,\,\,\kappa_p = |\pmb{\kappa}_p|, \nonumber\\
    &\mathbf{x} = \pmb{\kappa}_p+\pmb{\kappa}_b, \nonumber \\
    &x = \sqrt{\kappa_b^2+\kappa_p^2+2\kappa_b\kappa_p\cos(\phi_p-\phi_b)},\\
    &\phi_x = \phi_p\pm\arccos{\frac{\kappa_a^2+\kappa_p^2-\kappa_b^2}{2\kappa_a\kappa_p}}, \nonumber\\
    &\phi_b=\phi_p\pm\arccos{\frac{x^2-\kappa_b^2-\kappa_p^2}{2\kappa_b\kappa_p}}. \nonumber
\end{align}
The last integral in Eq.~\eqref{eq:TWM0} is evaluated with the help of the following identity
\begin{align}
\label{eq:dlint}
   \delta(\kappa_a-x)=\bigg[\frac{\delta(\phi_b-\phi_p-\delta_b)}{|\frac{\partial x}{\partial\phi_b}|}+\frac{\delta(\phi_b-\phi_{p}+\delta_b)}{|\frac{\partial x}{\partial\phi_b}|}\bigg].
\end{align}
To shorten the formula, we introduce notations for the phases 
\begin{align}
&\delta_b =\arccos{\frac{\kappa_a^2-\kappa_b^2-\kappa_p^2}{2\kappa_b\kappa_p}}, \nonumber \\
&\delta_x = \arccos{\frac{\kappa_a^2+\kappa_p^2-\kappa_b^2}{2\kappa_a\kappa_p}}, 
\end{align}
and the area of the triangle with the sides $\kappa_a,\kappa_b,\kappa_p$: 
\begin{align}
\label{eq:trian}
    \Delta = \frac{1}{4}\sqrt{4\kappa_b^2\kappa_p^2-(\kappa_a^2-\kappa_b^2-\kappa_p^2)^2}.
\end{align}
The resulting integral Eq.~\eqref{eq:TWM0} combined with Eq.~\eqref{eq:dlint} gives the following
\begin{widetext}
\begin{align}
    \langle\Phi_b^{\mathrm{TW}}|&e^{-i\mathbf{k}_p\mathbf{R}}|\Phi_a^{\mathrm{TW}}\rangle = i^{m_b-m_a}\cos\bigg[m_a\delta_x-m_b\delta_b\bigg]e^{i(m_a-m_b)\phi_p}\frac{\sqrt{\kappa_a\kappa_b}}{\Delta}
    \frac{\pi\delta(P_{z,a}-P_{z,b}-k_{z,p})}{2RL_z\sqrt{E_a E_b}}. 
\end{align}
Therefore, the $S$-matrix element equals
\begin{align}
\label{eq:TWC}
     S^{\rm TW} = 2\pi i S_e \delta(\varepsilon_a+E_a-\varepsilon_b-E_b-\omega)
    i^{m_b-m_a}\cos\bigg[m_a\delta_x-m_b\delta_b\bigg]e^{i(m_a-m_b)\phi_p}\frac{\sqrt{\kappa_a\kappa_b}}{\Delta}
    \frac{\pi\delta(P_{z,a}-P_{z,b}-k_{z,p})}{2RL_z\sqrt{E_a E_b}}\frac{1}{\sqrt{2\omega V}},
\end{align}
where $S_{e}$ is the electron matrix element given by Eq.~\eqref{eq:Spwe}.
The differential probability with Eq.~\eqref{eq:TWC} according to Eq.~\eqref{eq:prob} becomes
\begin{eqnarray}
     dw = \frac{|S_e|^2}{(2\pi)^3} \frac{1}{2E_a}\delta(\varepsilon_a+E_a-\varepsilon_b-E_b-\omega)\delta(P_{z,a}-P_{z,b}-k_{z,p})
    \bigg[1+\cos(2m_a\delta_x-2m_b\delta_b)\bigg]
     \frac{\kappa_b}{4\Delta}\frac{d^3 k_p}{2\omega}\frac{d\kappa_b\Delta m_b d P_{z,b}}{2E_b \pi},
\end{eqnarray}
\end{widetext}
where the final number of state for the twisted center of mass state is taken to be \cite{Jentschura:2011aa}
\begin{equation}
 dn_b = \frac{R d\kappa_b \Delta m_b}{\pi}\frac{L_z dP_{z,b}}{2\pi},
\end{equation}
and we utilized the following regularization of the $1/\Delta^2$ \cite{Ivanov2011}
\begin{align}
    \frac{1}{\Delta^2}=\frac{1}{\Delta}\frac{1}{2\pi}\int\limits_{0}^{2\pi} \frac{\delta(\kappa_a-x_\alpha)}{\kappa_a}d\alpha=\frac{1}{\Delta}\frac{R}{\pi \kappa_a}.
\end{align}
To proceed with the reduced probability and to perform the summation on the final center-of-mass state $\Phi_b$, it is convenient to represent the matrix element Eq.~\eqref{eq:melCTW} in a different form. We use the following identity for the exponent
\begin{align}
&e^{-i\mathbf{k}_p\pmb{R}} = e^{-i\kappa_p\rho\cos(\phi-\phi_p)}e^{-ik_{z,p}z}\nonumber\\&=e^{-ik_{z,p}z}\sum_{m_p} i^{m_p} J_{m_p}(\kappa_p \rho) e^{im_p(\phi_p-\phi+\pi)}
\end{align}
and compute the matrix center of mass matrix element 
\begin{align}
\label{eq:meTW}
&\langle\Phi_b^{\mathrm{TW}} | e^{-i\mathbf{k}_p\mathbf{R}} | \Phi_a^{\mathrm{TW}}\rangle = \frac{\pi}{RL_z} \sqrt{\frac{\kappa_a\kappa_b}{4 E_a E_b}} \nonumber \\
&\times \delta(P_{z,a}-P_{z,b}-k_{z,p}) e^{i(m_a-m_b)\phi_p}(-i)^{m_a-m_b} \nonumber \\
&\times\int\limits_0^\infty J_{m_b}(\kappa_b\rho)J_{m_a}(\kappa_a\rho)J_{m_a-m_b}(\kappa_p\rho)\rho d\rho.
\end{align}
The reduced probability can be found following Eq.\eqref{eq:probR} and reads
\begin{align}
    dw_r=\frac{|S_e|^2}{(2\pi)^3} \frac{d^3 k_p}{2\omega} \frac{1}{4E_a} \frac{\kappa_a}{R}\frac{\mathcal{I}_0}{2} 
\end{align}
Above we introduced the following notation
\begin{eqnarray}
     \mathcal{I}_0 &=& \sum_{m_b=-\infty}^\infty \int\limits_{0}^{\infty}\kappa_b d\kappa_b \frac{\delta(\varepsilon_a+E_a-\varepsilon_b-E_b-\omega)}{E_b}\nonumber\\
     &\times&\left|\int\limits_0^\infty J_{m_b}(\kappa_b\rho)J_{m_a}(\kappa_a\rho)J_{m_a-m_b}(\kappa_p\rho)\rho d\rho \right|^2.
\end{eqnarray}
We apply the regularization of the sum \cite{Ivanov2011}
\begin{align}
\label{eq:SumReg}
    \sum_{m_b=-\infty}^\infty &\left|\int\limits_0^\infty J_{m_b}(\kappa_b\rho)J_{m_a}(\kappa_a\rho)J_{m_a-m_b}(\kappa_p\rho)\rho d\rho \right|^2 \nonumber\\ &=\frac{R}{2\pi^2\kappa_a}\frac{1}{\Delta}
\end{align}
and arrive at
\begin{align}
\label{eq:redprob}
    dw_r=\frac{|S_e|^2}{(2\pi)^4} \frac{1}{8E_a} \frac{\mathcal{I}_1}{2 \pi} \omega d\omega d\Omega_p.
\end{align}
Above $\Delta$ is the area of the triangle with sides $\kappa_a, \kappa_b, \kappa_p$ and given by Eq.~\eqref{eq:trian}. The result is valid only if ($\kappa_a, \kappa_b, \kappa_p$) obeys the triangle inequality. Otherwise, the sum in Eq.~\eqref{eq:SumReg} is zero. Therefore, the final expression is nonzero only in the case of $|\kappa_a-\kappa_p|\leq\kappa_b\leq\kappa_a+\kappa_p$. The master integral $\mathcal{I}_1$ is given by
\begin{align}
\label{eq:MI1}
    \mathcal{I}_1=\int\limits_{|\kappa_a-\kappa_p|}^{\kappa_a+\kappa_p}\frac{4\delta(\varepsilon_a+E_a-\varepsilon_b-E_b-\omega)\kappa_b d\kappa_b}{E_b\sqrt{4\kappa_b^2\kappa_p^2-(\kappa_a^2-\kappa_b^2-\kappa_p^2)^2}}.
\end{align}
To check the consistency of the obtained result with the plane-wave case, we can consider the limiting case of $\kappa_a \rightarrow 0$. In this case we can use a substitution:
\begin{align}
\label{eq:subst}
    \frac{1}{2\pi \Delta}\Bigg|_{\kappa_a \rightarrow 0} = \frac{\delta(\kappa_b-\kappa_p)}{\kappa_b}.
\end{align}
and the master integral \eqref{eq:MI1} reads
\begin{eqnarray}
\label{eq:MI1lim}
\mathcal{I}_1\Big|_{\kappa_a \rightarrow 0} &=& 2 \pi \delta\left(\varepsilon_a-\varepsilon_b-\omega+\frac{P_{z,a}\omega\cos{\theta_p}}{M}-\frac{\omega^2}{2M}\right)\nonumber\\
&\times& \left(\frac{P_{z,a}^2}{2M}-\frac{P_{z,a}\omega\cos{\theta_p}}{M}+\frac{\omega^2}{2M}\right)^{-1}.
\end{eqnarray}
Substituting Eq.~\eqref{eq:MI1lim} into Eq.~\eqref{eq:redprob} we find an exact agreement with the plane wave result \eqref{eq:PredPLZ} up to a factor $(2\pi)^2$. The factor $(2\pi)^2$ is due to a different normalization of an initial state of the center of mass in the case of plane waves and twisted waves. 

The integral \eqref{eq:MI1} can be evaluated explicitly; however, the closed analytic solution is bulky. Note that $\kappa_b$ can be included in the differential. The argument of the delta function must vanish on the interval $\kappa_b^2\in \left[(\kappa_a-\kappa_p)^2, (\kappa_a+\kappa_p)^2 \right]$, otherwise the integral is zero. So we get
\begin{align}
\label{eq:I1ex}
  \mathcal{I}_1=\frac{4 M}{\tilde{E}_b\sqrt{4\tilde{\kappa_b}^2 \kappa_p^2-(\kappa_a^2-\tilde{\kappa_b}^2-\kappa_p^2)^2}}.  
\end{align}
Above 
\begin{align}
\label{eq:msIntE}
    \tilde{E}_b=\frac{\tilde{\kappa_b}^2}{2M}+\frac{(P_{z,a}-\omega \cos \theta_p)^2}{2M},
\end{align}
and 
\begin{align}
    \tilde{\kappa}_b^2=2M(\varepsilon_a-\varepsilon_b-\omega)+P_a^2-(P_{z,a}-\omega\cos\theta_p)^2.
\end{align}

The equation \eqref{eq:msIntE} has two discontinuities at the points $\theta_p=\theta_p^{PW}\pm\theta_a$, where $\theta_p^{PW}$ is the angle of maximum intensity of the emitted photons in the plane wave case.

In order to get a quantitative understanding and to compare Eq.\eqref{eq:msIntE} and Eq.\eqref{eq:MI1} with the plane-wave case Eq.\eqref{eq:PredPLZ} we first note that the electron matrix element $|S_e|^2$ is the same and we can drop it in the comparison. Next, we restrict our comparison to the structure of the distribution only, so we eliminate the exact normalization factors and normalize all subsequent results to their maximum values.

\begin{figure}[t]
    \includegraphics[width=0.47\textwidth]{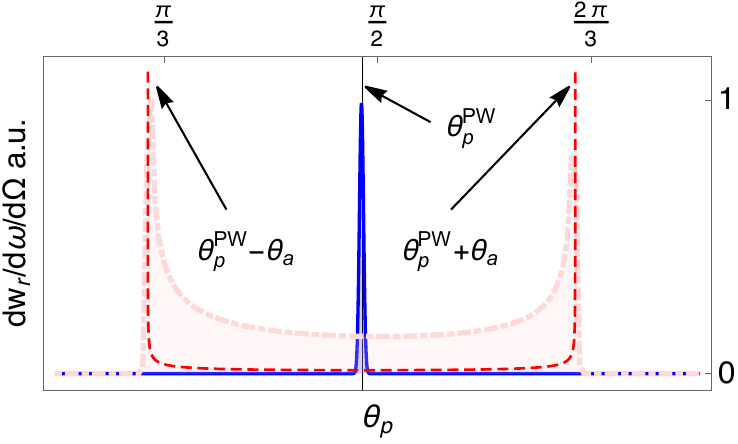}
    \caption{Differential photon density distributions $dw_r/d\omega/d\Omega $ normalised to the corresponding maximum value for the two cases: solid blue line - plane wave case given by Eq.\eqref{eq:PredPLZ} with delta function replaced according to the Eq.\eqref{eq:dlG}; dot dashed light red line - twisted wave case Eq.\eqref{eq:redprob} with the $\mathcal{I}_1$ Eq.\eqref{eq:MI1} calculated numerically with the substitution Eq.\eqref{eq:dlG}; red dashed line - twisted wave case Eq.\eqref{eq:redprob} with the $\mathcal{I}_1$ calculated exactly and given by Eq.\eqref{eq:I1ex}.  To produce the plots we have used synthetic parameters $P_a=1$, $k_p=0.1$, $M=1$, $\varepsilon_a-\varepsilon_b-\omega=10^{-3}$, $\theta_a=\pi/6$, $\sigma_e=5\times10^{-4}$. \label{Fig:N}}
\end{figure}

We regularize the energy delta function by replacing it with a narrow Gaussian distribution with an effective $\sigma_E$ that is small but finite. 
\begin{align}
\label{eq:dlG}
 \delta(E)\to \frac{1}{\sqrt{2\pi}\sigma_E}\exp \left( -\frac{E^2}{2\sigma_E^2} \right).   
\end{align}
This allows us to evaluate all expressions numerically.      

In Fig.\ref{Fig:N} we plot $\frac{dw_r}{d\omega d\Omega}$ given by Eq.\eqref{eq:PredPLZ} and normalized to it's maximum value and compare it with the same quantity derived from Eq.\eqref{eq:redprob}. We show two different cases: first, the numerical evaluation of the integral \eqref{eq:MI1} with the replacement \eqref{eq:dlG} and second, the exact where the master integral is given by Eq.\eqref{eq:msIntE}.

We observe that for reasonably large opening angles of the center of mass state $\theta_a$ the photon distribution is modified and split into two peaks which are symmetrical with respect to the intensity peak of the plane wave case $\theta_p^{PW}$. We note that the angular shift is exactly $\pm \theta_a$ with respect to $\theta_p^{PW}$. and can be observed whenever the opening angle $\theta_a$ is not small.

We conclude that the distribution of the emitted photons of the atomic system with the twisted center of mass state differs from the common plane wave case. Such a difference may be observed in the experiment.

\section{Twisted to plane wave}


Now we consider the case when the initial state of an atom is a twisted wave and the final state is given by a plane wave. This scenario corresponds to the detection of the atom in the final state with the help of a common detector that allows one to measure the intensity of the atomic flux under a fixed angle with respect to the propagation axis of the initial twisted atomic beam. Therefore, the initial state of the center of mass in this configuration is a twisted wave given by Eq.~\eqref{eq:TWW} and the final state is a plane wave \eqref{eq:PWF}. In the lowest order in $m/M$ with the interaction Hamiltonian given by Eq.~\eqref{eq:Hint} the $S$ matrix is then
\begin{eqnarray}
\label{eq:SMTP}    
    S^{\rm PWTW} &=& \frac{2\pi i}{\sqrt{2\omega V}} S_e \delta(\varepsilon_a+E_a-\varepsilon_b-E_b-\omega) \nonumber\\
    &\times&\langle \Phi_b^{\rm PW}|e^{-i\mathbf{k}_p\mathbf{R}}| \Phi_a^{\rm TW} \rangle.
\end{eqnarray}
The center-of-mass matrix element is evaluated using the representation Eq.~\eqref{eq:TWW2} for the twisted wave and can be expressed as
\begin{align}
    &\langle \Phi_b^{\rm PW}|e^{-i\mathbf{k}_p\mathbf{R}}| \Phi_a^{\rm TW} \rangle = (2\pi)^{3/2}\sqrt{\frac{\kappa_a}{4E_aE_b}}\sqrt{\frac{\pi}{RL_zV}} \nonumber\\
    &\times\int \delta^3(\mathbf{P}_a-\mathbf{k}_p-\mathbf{P}_b)i^{-m_a}e^{im_a\phi_a}d\phi_a.
\end{align}
Representing the $\delta$ function in cylindrical coordinates, we compute the integral and get
\begin{align}
\label{eq:CMPT}
    &\langle \Phi_b^{\rm PW}|e^{-i\mathbf{k}_p\mathbf{R}}| \Phi_a^{\rm TW} \rangle = (2\pi)^{3/2}\sqrt{\frac{1}{4E_aE_b\kappa_a}}\sqrt{\frac{\pi}{RL_zV}}  \nonumber\\
    &\times i^{-m_a}e^{im_a\phi_x}\delta(P_{z,a}-P_{z,b}-k_{z,p})\delta(\kappa_a-x_0).
\end{align}
Here along with notations Eq.~\eqref{eq:not} we introduced
\begin{align}
     &\phi_{x_0}=\phi_b+ \angle (\mathbf{x}_0,\boldsymbol{\kappa}_b), \nonumber\\
    &\mathbf{x}_0=\boldsymbol{\kappa}_p+\boldsymbol{\kappa}_b, \\   &x_0=|\mathbf{x}_0|=\sqrt{\kappa_b^2+\kappa_p^2+2\kappa_b\kappa_p\cos(\phi_p-\phi_b)}.\nonumber
\end{align}
With Eq.~\eqref{eq:SMTP} and Eq.~\eqref{eq:CMPT} expression for the $S$-matrix reads
\begin{eqnarray}
    S^{\rm PWTW} &= &\frac{(2\pi)^{5/2}i}{\sqrt{2\omega V}}\sqrt{\frac{1}{4E_aE_b\kappa_a}}\sqrt{\frac{\pi}{RL_zV}}S_e\nonumber\\
    &\times& \delta(\varepsilon_a+E_a-\varepsilon_b-E_b-\omega)\nonumber\\
    &\times& i^{-m_a}e^{im_a\phi_x}\delta(P_{z,a}-P_{z,b}-k_{z,p})\nonumber\\
    &\times&\delta(\kappa_a-x_0).
\end{eqnarray}
With the help of the following regularization
\begin{align}
    &\left[\delta(P_{z,a}-P_{z,b}-k_{z,p}) \right]^2 = \frac{L_z}{2\pi}\delta(P_{z,a}-P_{z,b}-k_{z,p}), \nonumber\\
    &\left[\delta(\kappa_{a}-x_0) \right]^2 = \frac{R}{\pi}\delta(\kappa_{a}-x_0).
\end{align}
we finally get for the differential probability
\begin{align}
    &dw = \frac{|S_e|^2}{(2\pi)^3}\frac{1}{2E_a}\delta(\varepsilon_a+E_a-\varepsilon_b-E_b-\omega)\nonumber\\
    &\times \delta(P_{z,a}-P_{z,b}-k_{z,p})\frac{\delta(\kappa_{a}-x_0)}{\kappa_a}\frac{ d^3 P_b}{2E_b}\frac{ d^3 k_p}{2\omega}.
\end{align}
\begin{figure}[t]
    \includegraphics[width=0.4\textwidth]{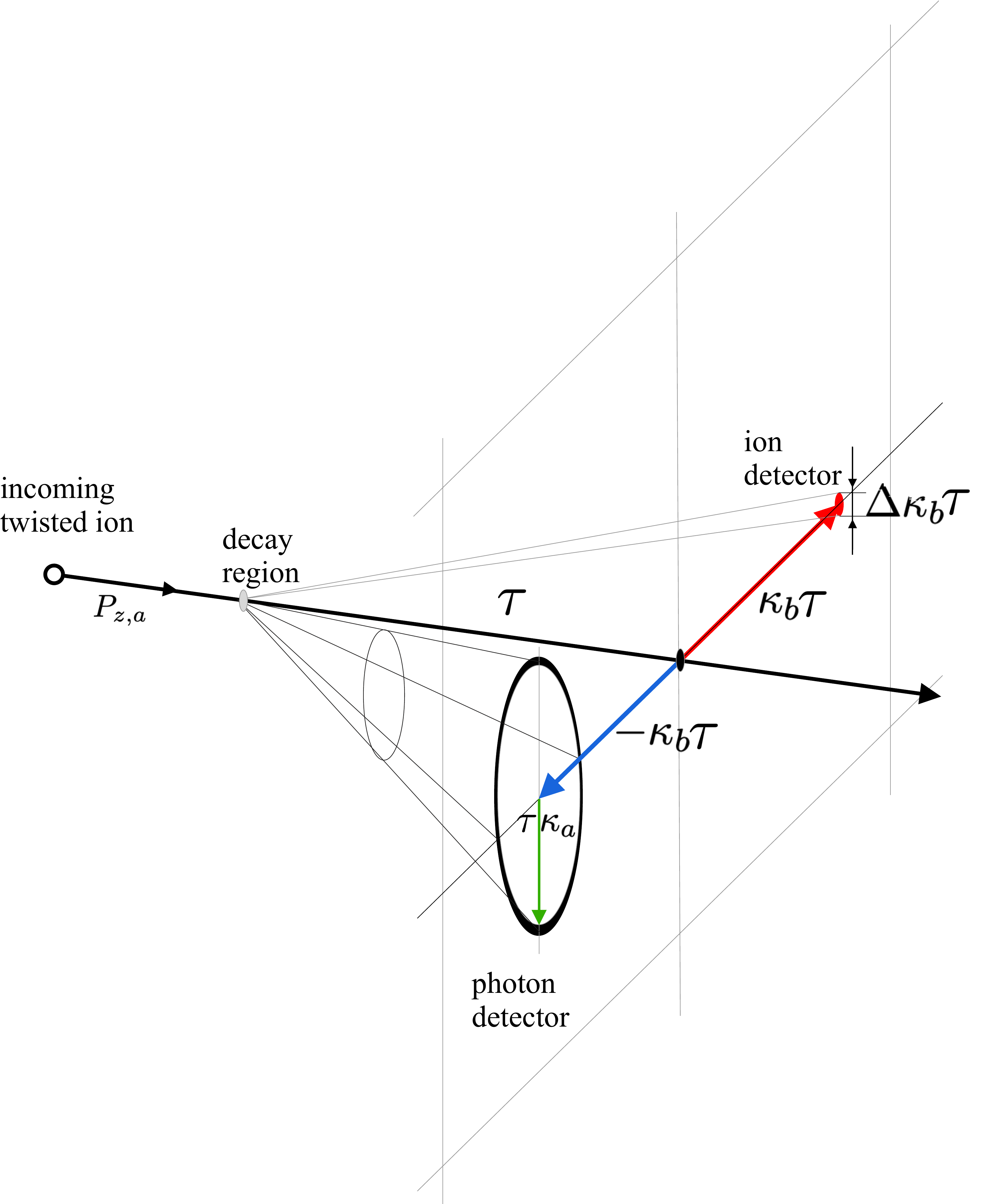}
    \caption{Sketch of the coincidence experiment for the simultaneous detection of the final state of the atom and the photon distribution in momentum space. Once the transverse momentum of the atom $\kappa_b$ is fixed, the transverse momentum of the photon is restricted to the circle with radius $\kappa_a$ and displacement from the origin $-\kappa_b$ as dictated by Eq.~\eqref{eq:trmomFin}. In the sketch above we have assumed that the time of flight $\tau$ from the decay point to the observation plane is known and that the characteristic decay time is significantly less than the time of flight.\label{Fig:2}}
\end{figure}
When using a coincidence circuit to detect both the final state of the atom and the emitted photon, the radial delta function limits the photon emission angles in the plane perpendicular to the initial axis of atom propagation (see Fig.~\ref{Fig:2} for details). For an atom detector of small angular size, intercepting atoms with final transverse momentum $\Delta \kappa_b$ when the condition $\Delta \kappa_b/ \kappa_b\ll 1$ is satisfied, the reduced probability (an integral over $d^3 P_b$) is
\begin{eqnarray}
\label{eq:redprobpl}
    dw_r &\approx& \frac{|S_e|^2}{(2\pi)^2}\frac{ d^3 k_p}{2\omega}\delta(\varepsilon_a+E_a-\varepsilon_b-E_b-\omega) \nonumber\\ &\times&\frac{1}{2E_a} \frac{L_z}{2\pi}\frac{R}{\pi}\frac{\delta(\kappa_{a}-x_0)}{\kappa_a} \frac{2\pi \kappa_b \Delta \kappa_b}{2E_b}.
\end{eqnarray}
For the cases $\kappa_b > \kappa_a$ and $\kappa_b \sim \kappa_a$ in the coincidence circuit detector, the probability is nonzero only if the argument of the radial delta function vanishes. This corresponds to the following connection between the transverse components of the momentum.
\begin{align}
\label{eq:trmom}
    \kappa_b^2+\kappa_p^2+2\kappa_b\kappa_p\cos\phi_p-\kappa_a^2=0.
\end{align}
where, without loss of generality, we set $\phi_b = 0$.

For fixed values of $\kappa_a$ and $\kappa_b$ the equation Eq.~\eqref{eq:trmom} is an equation of the displaced circle in the transverse plane of the momentum space for the transverse part of the photon wave vector $\boldsymbol{\kappa}_p$.
\begin{align}
\label{eq:trmomFin}
    (\kappa_x+\kappa_b)^2 + \kappa_y^2 = \kappa_a^2.
\end{align}
where $\boldsymbol{\kappa}_p = \kappa_x \mathbf{e}_x+\kappa_y \mathbf{e}_y$. 

We immediately observe that if atoms are detected in a small region of the momentum space, then the corresponding photons resemble a ring in the transverse momentum plane with the center at the point $\kappa_x = -\kappa_b$ and radius $\kappa_a$. We note that despite the fact that the photon distribution resembles a ring, no conclusions can be drawn about the phase of the photons and their OAM. Thus, the proposed experimental setup only reveals the twisted nature of the initial atomic state.    


\section{Conclusion}


We have introduced a model of a twisted atom based on the quantum field description and the $S$ matrix formalism. Within the formalism, we introduced the center of mass and the relative (electron) coordinates, which allows us to reduce the full Hamiltonian to a Schr\"odinger Hamiltonian for the free center of mass and a Coulomb Hamiltonian for the bound electron. By finding the solution of the free Schr\"odinger equation for the center of mass in cylindrical coordinates, we have arrived at a vortex atomic state. Furthermore, we have studied the influence of the center of mass quantum state on the properties of the photons emitted during the electron transitions. We have studied the influence of the initially twisted center of mass state in the lowest order of the electron-nucleus mass ratio. We have shown that in a common scenario where the final state of the atom is not detected, the angular distribution of the emitted photons is altered.  The latter follows from Eq.\eqref{eq:redprob}, which together with Eq.\eqref{eq:redprobpl} are the main results of the present investigation.
Finally, we conclude that in a specially arranged coincidence scheme, the initial twist of the center of mass can be confirmed by measuring the intensity distribution of the emitted photons.

\begin{acknowledgments}
The authors thank Igor Chestnov, Dmitriy Karlovets, and Ivan Terekhov for useful discussions and suggestions.
\end{acknowledgments}

\bibliography{refs}

\begin{thebibliography}{51}%
\makeatletter
\providecommand \@ifxundefined [1]{%
 \@ifx{#1\undefined}
}%
\providecommand \@ifnum [1]{%
 \ifnum #1\expandafter \@firstoftwo
 \else \expandafter \@secondoftwo
 \fi
}%
\providecommand \@ifx [1]{%
 \ifx #1\expandafter \@firstoftwo
 \else \expandafter \@secondoftwo
 \fi
}%
\providecommand \natexlab [1]{#1}%
\providecommand \enquote  [1]{``#1''}%
\providecommand \bibnamefont  [1]{#1}%
\providecommand \bibfnamefont [1]{#1}%
\providecommand \citenamefont [1]{#1}%
\providecommand \href@noop [0]{\@secondoftwo}%
\providecommand \href [0]{\begingroup \@sanitize@url \@href}%
\providecommand \@href[1]{\@@startlink{#1}\@@href}%
\providecommand \@@href[1]{\endgroup#1\@@endlink}%
\providecommand \@sanitize@url [0]{\catcode `\\12\catcode `\$12\catcode
  `\&12\catcode `\#12\catcode `\^12\catcode `\_12\catcode `\%12\relax}%
\providecommand \@@startlink[1]{}%
\providecommand \@@endlink[0]{}%
\providecommand \url  [0]{\begingroup\@sanitize@url \@url }%
\providecommand \@url [1]{\endgroup\@href {#1}{\urlprefix }}%
\providecommand \urlprefix  [0]{URL }%
\providecommand \Eprint [0]{\href }%
\providecommand \doibase [0]{https://doi.org/}%
\providecommand \selectlanguage [0]{\@gobble}%
\providecommand \bibinfo  [0]{\@secondoftwo}%
\providecommand \bibfield  [0]{\@secondoftwo}%
\providecommand \translation [1]{[#1]}%
\providecommand \BibitemOpen [0]{}%
\providecommand \bibitemStop [0]{}%
\providecommand \bibitemNoStop [0]{.\EOS\space}%
\providecommand \EOS [0]{\spacefactor3000\relax}%
\providecommand \BibitemShut  [1]{\csname bibitem#1\endcsname}%
\let\auto@bib@innerbib\@empty
\bibitem [{\citenamefont {Allen}\ \emph {et~al.}(1992)\citenamefont {Allen},
  \citenamefont {Beijersbergen}, \citenamefont {Spreeuw},\ and\ \citenamefont
  {Woerdman}}]{Allen1}%
  \BibitemOpen
  \bibfield  {author} {\bibinfo {author} {\bibfnamefont {L.}~\bibnamefont
  {Allen}}, \bibinfo {author} {\bibfnamefont {M.~W.}\ \bibnamefont
  {Beijersbergen}}, \bibinfo {author} {\bibfnamefont {R.~J.~C.}\ \bibnamefont
  {Spreeuw}},\ and\ \bibinfo {author} {\bibfnamefont {J.~P.}\ \bibnamefont
  {Woerdman}},\ }\href {https://doi.org/10.1103/PhysRevA.45.8185} {\bibfield
  {journal} {\bibinfo  {journal} {Phys. Rev. A}\ }\textbf {\bibinfo {volume}
  {45}},\ \bibinfo {pages} {8185} (\bibinfo {year} {1992})}\BibitemShut
  {NoStop}%
\bibitem [{\citenamefont {Allen}\ \emph {et~al.}(1999)\citenamefont {Allen},
  \citenamefont {Padgett},\ and\ \citenamefont {Babiker}}]{Allen1999}%
  \BibitemOpen
  \bibfield  {author} {\bibinfo {author} {\bibfnamefont {L.}~\bibnamefont
  {Allen}}, \bibinfo {author} {\bibfnamefont {M.}~\bibnamefont {Padgett}},\
  and\ \bibinfo {author} {\bibfnamefont {M.}~\bibnamefont {Babiker}}\
  }(\bibinfo  {publisher} {Elsevier},\ \bibinfo {year} {1999})\ pp.\ \bibinfo
  {pages} {291--372}\BibitemShut {NoStop}%
\bibitem [{\citenamefont {Franke-Arnold}\ \emph {et~al.}(2008)\citenamefont
  {Franke-Arnold}, \citenamefont {Allen},\ and\ \citenamefont
  {Padgett}}]{FrankeArnold2008}%
  \BibitemOpen
  \bibfield  {author} {\bibinfo {author} {\bibfnamefont {S.}~\bibnamefont
  {Franke-Arnold}}, \bibinfo {author} {\bibfnamefont {L.}~\bibnamefont
  {Allen}},\ and\ \bibinfo {author} {\bibfnamefont {M.}~\bibnamefont
  {Padgett}},\ }\href {https://doi.org/https://doi.org/10.1002/lpor.200810007}
  {\bibfield  {journal} {\bibinfo  {journal} {Laser \& Photonics Reviews}\
  }\textbf {\bibinfo {volume} {2}},\ \bibinfo {pages} {299} (\bibinfo {year}
  {2008})}\BibitemShut {NoStop}%
\bibitem [{\citenamefont {Andrews}(2012)}]{ct3}%
  \BibitemOpen
  \bibinfo {editor} {\bibfnamefont {D.~L.}\ \bibnamefont {Andrews}},\ ed.,\
  \href {https://doi.org/10.1017/CBO9780511795213} {\emph {\bibinfo {title}
  {The Angular Momentum of Light}}}\ (\bibinfo  {publisher} {Cambridge
  University Press},\ \bibinfo {year} {2012})\BibitemShut {NoStop}%
\bibitem [{\citenamefont {Padgett}\ \emph {et~al.}(2004)\citenamefont
  {Padgett}, \citenamefont {Courtial},\ and\ \citenamefont {Allen}}]{TL2}%
  \BibitemOpen
  \bibfield  {author} {\bibinfo {author} {\bibfnamefont {M.}~\bibnamefont
  {Padgett}}, \bibinfo {author} {\bibfnamefont {J.}~\bibnamefont {Courtial}},\
  and\ \bibinfo {author} {\bibfnamefont {L.}~\bibnamefont {Allen}},\ }\href
  {https://doi.org/10.1063/1.1768672} {\bibfield  {journal} {\bibinfo
  {journal} {Physics Today}\ }\textbf {\bibinfo {volume} {57}},\ \bibinfo
  {pages} {35} (\bibinfo {year} {2004})}\BibitemShut {NoStop}%
\bibitem [{\citenamefont {Yao}\ and\ \citenamefont {Padgett}(2011)}]{TW3}%
  \BibitemOpen
  \bibfield  {author} {\bibinfo {author} {\bibfnamefont {A.~M.}\ \bibnamefont
  {Yao}}\ and\ \bibinfo {author} {\bibfnamefont {M.~J.}\ \bibnamefont
  {Padgett}},\ }\href {https://doi.org/10.1364/AOP.3.000161} {\bibfield
  {journal} {\bibinfo  {journal} {Adv. Opt. Photon.}\ }\textbf {\bibinfo
  {volume} {3}},\ \bibinfo {pages} {161} (\bibinfo {year} {2011})}\BibitemShut
  {NoStop}%
\bibitem [{\citenamefont {Bahrdt}\ \emph {et~al.}(2013)\citenamefont {Bahrdt},
  \citenamefont {Holldack}, \citenamefont {Kuske}, \citenamefont {M\"uller},
  \citenamefont {Scheer},\ and\ \citenamefont {Schmid}}]{TL4}%
  \BibitemOpen
  \bibfield  {author} {\bibinfo {author} {\bibfnamefont {J.}~\bibnamefont
  {Bahrdt}}, \bibinfo {author} {\bibfnamefont {K.}~\bibnamefont {Holldack}},
  \bibinfo {author} {\bibfnamefont {P.}~\bibnamefont {Kuske}}, \bibinfo
  {author} {\bibfnamefont {R.}~\bibnamefont {M\"uller}}, \bibinfo {author}
  {\bibfnamefont {M.}~\bibnamefont {Scheer}},\ and\ \bibinfo {author}
  {\bibfnamefont {P.}~\bibnamefont {Schmid}},\ }\href
  {https://doi.org/10.1103/PhysRevLett.111.034801} {\bibfield  {journal}
  {\bibinfo  {journal} {Phys. Rev. Lett.}\ }\textbf {\bibinfo {volume} {111}},\
  \bibinfo {pages} {034801} (\bibinfo {year} {2013})}\BibitemShut {NoStop}%
\bibitem [{\citenamefont {Knyazev}\ and\ \citenamefont {Serbo}(2018)}]{UFN}%
  \BibitemOpen
  \bibfield  {author} {\bibinfo {author} {\bibfnamefont {B.}~\bibnamefont
  {Knyazev}}\ and\ \bibinfo {author} {\bibfnamefont {V.}~\bibnamefont
  {Serbo}},\ }\href@noop {} {\bibfield  {journal} {\bibinfo  {journal}
  {Phys.-Usp.}\ }\textbf {\bibinfo {volume} {61}},\ \bibinfo {pages} {449}
  (\bibinfo {year} {2018})}\BibitemShut {NoStop}%
\bibitem [{\citenamefont {Bliokh}\ \emph {et~al.}(2007)\citenamefont {Bliokh},
  \citenamefont {Bliokh}, \citenamefont {Savel'ev},\ and\ \citenamefont
  {Nori}}]{Bliokh2007}%
  \BibitemOpen
  \bibfield  {author} {\bibinfo {author} {\bibfnamefont {K.~Y.}\ \bibnamefont
  {Bliokh}}, \bibinfo {author} {\bibfnamefont {Y.~P.}\ \bibnamefont {Bliokh}},
  \bibinfo {author} {\bibfnamefont {S.}~\bibnamefont {Savel'ev}},\ and\
  \bibinfo {author} {\bibfnamefont {F.}~\bibnamefont {Nori}},\ }\href
  {https://doi.org/10.1103/PhysRevLett.99.190404} {\bibfield  {journal}
  {\bibinfo  {journal} {Phys. Rev. Lett.}\ }\textbf {\bibinfo {volume} {99}},\
  \bibinfo {pages} {190404} (\bibinfo {year} {2007})}\BibitemShut {NoStop}%
\bibitem [{\citenamefont {Uchida}(2010)}]{ushida2010}%
  \BibitemOpen
  \bibfield  {author} {\bibinfo {author} {\bibfnamefont {A.}~\bibnamefont
  {Uchida}, \bibfnamefont {M.~Tonomura}},\ }\href
  {https://doi.org/doi.org/10.1038/nature08904} {\bibfield  {journal} {\bibinfo
   {journal} {Nature}\ }\textbf {\bibinfo {volume} {464}},\ \bibinfo {pages}
  {737} (\bibinfo {year} {2010})}\BibitemShut {NoStop}%
\bibitem [{\citenamefont {McMorran}\ \emph {et~al.}(2011)\citenamefont
  {McMorran}, \citenamefont {Agrawal}, \citenamefont {Anderson}, \citenamefont
  {Herzing}, \citenamefont {Lezec}, \citenamefont {McClelland},\ and\
  \citenamefont {Unguris}}]{McMorran2011}%
  \BibitemOpen
  \bibfield  {author} {\bibinfo {author} {\bibfnamefont {B.~J.}\ \bibnamefont
  {McMorran}}, \bibinfo {author} {\bibfnamefont {A.}~\bibnamefont {Agrawal}},
  \bibinfo {author} {\bibfnamefont {I.~M.}\ \bibnamefont {Anderson}}, \bibinfo
  {author} {\bibfnamefont {A.~A.}\ \bibnamefont {Herzing}}, \bibinfo {author}
  {\bibfnamefont {H.~J.}\ \bibnamefont {Lezec}}, \bibinfo {author}
  {\bibfnamefont {J.~J.}\ \bibnamefont {McClelland}},\ and\ \bibinfo {author}
  {\bibfnamefont {J.}~\bibnamefont {Unguris}},\ }\href
  {https://doi.org/10.1126/science.1198804} {\bibfield  {journal} {\bibinfo
  {journal} {Science}\ }\textbf {\bibinfo {volume} {331}},\ \bibinfo {pages}
  {192} (\bibinfo {year} {2011})}\BibitemShut {NoStop}%
\bibitem [{\citenamefont {Bliokh}\ \emph {et~al.}(2012)\citenamefont {Bliokh},
  \citenamefont {Schattschneider}, \citenamefont {Verbeeck},\ and\
  \citenamefont {Nori}}]{Bliokh2012}%
  \BibitemOpen
  \bibfield  {author} {\bibinfo {author} {\bibfnamefont {K.~Y.}\ \bibnamefont
  {Bliokh}}, \bibinfo {author} {\bibfnamefont {P.}~\bibnamefont
  {Schattschneider}}, \bibinfo {author} {\bibfnamefont {J.}~\bibnamefont
  {Verbeeck}},\ and\ \bibinfo {author} {\bibfnamefont {F.}~\bibnamefont
  {Nori}},\ }\href {https://doi.org/10.1103/PhysRevX.2.041011} {\bibfield
  {journal} {\bibinfo  {journal} {Phys. Rev. X}\ }\textbf {\bibinfo {volume}
  {2}},\ \bibinfo {pages} {041011} (\bibinfo {year} {2012})}\BibitemShut
  {NoStop}%
\bibitem [{\citenamefont {Clark}\ \emph {et~al.}(2015)\citenamefont {Clark},
  \citenamefont {Barankov}, \citenamefont {Huber}, \citenamefont {Arif},
  \citenamefont {Cory},\ and\ \citenamefont {Pushin}}]{ct8}%
  \BibitemOpen
  \bibfield  {author} {\bibinfo {author} {\bibfnamefont {C.~W.}\ \bibnamefont
  {Clark}}, \bibinfo {author} {\bibfnamefont {R.}~\bibnamefont {Barankov}},
  \bibinfo {author} {\bibfnamefont {M.~G.}\ \bibnamefont {Huber}}, \bibinfo
  {author} {\bibfnamefont {M.}~\bibnamefont {Arif}}, \bibinfo {author}
  {\bibfnamefont {D.~G.}\ \bibnamefont {Cory}},\ and\ \bibinfo {author}
  {\bibfnamefont {D.~A.}\ \bibnamefont {Pushin}},\ }\href
  {https://doi.org/10.1038/nature15265} {\bibfield  {journal} {\bibinfo
  {journal} {Nature}\ }\textbf {\bibinfo {volume} {525}},\ \bibinfo {pages}
  {504} (\bibinfo {year} {2015})}\BibitemShut {NoStop}%
\bibitem [{\citenamefont {Sarenac}\ \emph {et~al.}(2018)\citenamefont
  {Sarenac}, \citenamefont {Nsofini}, \citenamefont {Hincks}, \citenamefont
  {Arif}, \citenamefont {Clark}, \citenamefont {Cory}, \citenamefont {Huber},\
  and\ \citenamefont {Pushin}}]{Sarenac_2018}%
  \BibitemOpen
  \bibfield  {author} {\bibinfo {author} {\bibfnamefont {D.}~\bibnamefont
  {Sarenac}}, \bibinfo {author} {\bibfnamefont {J.}~\bibnamefont {Nsofini}},
  \bibinfo {author} {\bibfnamefont {I.}~\bibnamefont {Hincks}}, \bibinfo
  {author} {\bibfnamefont {M.}~\bibnamefont {Arif}}, \bibinfo {author}
  {\bibfnamefont {C.~W.}\ \bibnamefont {Clark}}, \bibinfo {author}
  {\bibfnamefont {D.~G.}\ \bibnamefont {Cory}}, \bibinfo {author}
  {\bibfnamefont {M.~G.}\ \bibnamefont {Huber}},\ and\ \bibinfo {author}
  {\bibfnamefont {D.~A.}\ \bibnamefont {Pushin}},\ }\href
  {https://doi.org/10.1088/1367-2630/aae3ac} {\bibfield  {journal} {\bibinfo
  {journal} {New Journal of Physics}\ }\textbf {\bibinfo {volume} {20}},\
  \bibinfo {pages} {103012} (\bibinfo {year} {2018})}\BibitemShut {NoStop}%
\bibitem [{\citenamefont {Jach}\ and\ \citenamefont {Vinson}(2022)}]{jach2022}%
  \BibitemOpen
  \bibfield  {author} {\bibinfo {author} {\bibfnamefont {T.}~\bibnamefont
  {Jach}}\ and\ \bibinfo {author} {\bibfnamefont {J.}~\bibnamefont {Vinson}},\
  }\href {https://doi.org/10.1103/PhysRevC.105.L061601} {\bibfield  {journal}
  {\bibinfo  {journal} {Phys. Rev. C}\ }\textbf {\bibinfo {volume} {105}},\
  \bibinfo {pages} {L061601} (\bibinfo {year} {2022})}\BibitemShut {NoStop}%
\bibitem [{\citenamefont {Lembessis}\ \emph {et~al.}(2014)\citenamefont
  {Lembessis}, \citenamefont {Ellinas}, \citenamefont {Babiker},\ and\
  \citenamefont {Al-Dossary}}]{Lembessis2014}%
  \BibitemOpen
  \bibfield  {author} {\bibinfo {author} {\bibfnamefont {V.~E.}\ \bibnamefont
  {Lembessis}}, \bibinfo {author} {\bibfnamefont {D.}~\bibnamefont {Ellinas}},
  \bibinfo {author} {\bibfnamefont {M.}~\bibnamefont {Babiker}},\ and\ \bibinfo
  {author} {\bibfnamefont {O.}~\bibnamefont {Al-Dossary}},\ }\href
  {https://doi.org/10.1103/PhysRevA.89.053616} {\bibfield  {journal} {\bibinfo
  {journal} {Phys. Rev. A}\ }\textbf {\bibinfo {volume} {89}},\ \bibinfo
  {pages} {053616} (\bibinfo {year} {2014})}\BibitemShut {NoStop}%
\bibitem [{\citenamefont {Hayrapetyan}\ and\ \citenamefont
  {Fritzsche}(2013)}]{Hayrapetyan_2013}%
  \BibitemOpen
  \bibfield  {author} {\bibinfo {author} {\bibfnamefont {A.~G.}\ \bibnamefont
  {Hayrapetyan}}\ and\ \bibinfo {author} {\bibfnamefont {S.}~\bibnamefont
  {Fritzsche}},\ }\href {https://doi.org/10.1088/0031-8949/2013/T156/014067}
  {\bibfield  {journal} {\bibinfo  {journal} {Physica Scripta}\ }\textbf
  {\bibinfo {volume} {2013}},\ \bibinfo {pages} {014067} (\bibinfo {year}
  {2013})}\BibitemShut {NoStop}%
\bibitem [{\citenamefont {Hayrapetyan}\ \emph {et~al.}(2013)\citenamefont
  {Hayrapetyan}, \citenamefont {Matula}, \citenamefont {Surzhykov},\ and\
  \citenamefont {Fritzsche}}]{Har2}%
  \BibitemOpen
  \bibfield  {author} {\bibinfo {author} {\bibfnamefont {A.~G.}\ \bibnamefont
  {Hayrapetyan}}, \bibinfo {author} {\bibfnamefont {O.}~\bibnamefont {Matula}},
  \bibinfo {author} {\bibfnamefont {A.}~\bibnamefont {Surzhykov}},\ and\
  \bibinfo {author} {\bibfnamefont {S.}~\bibnamefont {Fritzsche}},\ }\href
  {https://doi.org/10.1140/epjd/e2013-30191-x} {\bibfield  {journal} {\bibinfo
  {journal} {The European Physical Journal D}\ }\textbf {\bibinfo {volume}
  {67}},\ \bibinfo {pages} {167} (\bibinfo {year} {2013})}\BibitemShut
  {NoStop}%
\bibitem [{\citenamefont {Matula}\ \emph {et~al.}(2013)\citenamefont {Matula},
  \citenamefont {Hayrapetyan}, \citenamefont {Serbo}, \citenamefont
  {Surzhykov},\ and\ \citenamefont {Fritzsche}}]{Matula_2013}%
  \BibitemOpen
  \bibfield  {author} {\bibinfo {author} {\bibfnamefont {O.}~\bibnamefont
  {Matula}}, \bibinfo {author} {\bibfnamefont {A.~G.}\ \bibnamefont
  {Hayrapetyan}}, \bibinfo {author} {\bibfnamefont {V.~G.}\ \bibnamefont
  {Serbo}}, \bibinfo {author} {\bibfnamefont {A.}~\bibnamefont {Surzhykov}},\
  and\ \bibinfo {author} {\bibfnamefont {S.}~\bibnamefont {Fritzsche}},\ }\href
  {https://doi.org/10.1088/0953-4075/46/20/205002} {\bibfield  {journal}
  {\bibinfo  {journal} {Journal of Physics B: Atomic, Molecular and Optical
  Physics}\ }\textbf {\bibinfo {volume} {46}},\ \bibinfo {pages} {205002}
  (\bibinfo {year} {2013})}\BibitemShut {NoStop}%
\bibitem [{\citenamefont {Surzhykov}\ \emph {et~al.}(2015)\citenamefont
  {Surzhykov}, \citenamefont {Seipt}, \citenamefont {Serbo},\ and\
  \citenamefont {Fritzsche}}]{surzhykov2015}%
  \BibitemOpen
  \bibfield  {author} {\bibinfo {author} {\bibfnamefont {A.}~\bibnamefont
  {Surzhykov}}, \bibinfo {author} {\bibfnamefont {D.}~\bibnamefont {Seipt}},
  \bibinfo {author} {\bibfnamefont {V.~G.}\ \bibnamefont {Serbo}},\ and\
  \bibinfo {author} {\bibfnamefont {S.}~\bibnamefont {Fritzsche}},\ }\href
  {https://doi.org/10.1103/PhysRevA.91.013403} {\bibfield  {journal} {\bibinfo
  {journal} {Phys. Rev. A}\ }\textbf {\bibinfo {volume} {91}},\ \bibinfo
  {pages} {013403} (\bibinfo {year} {2015})}\BibitemShut {NoStop}%
\bibitem [{\citenamefont {Surzhykov}\ \emph {et~al.}(2016)\citenamefont
  {Surzhykov}, \citenamefont {Seipt},\ and\ \citenamefont
  {Fritzsche}}]{surzhykov:2016:033420}%
  \BibitemOpen
  \bibfield  {author} {\bibinfo {author} {\bibfnamefont {A.}~\bibnamefont
  {Surzhykov}}, \bibinfo {author} {\bibfnamefont {D.}~\bibnamefont {Seipt}},\
  and\ \bibinfo {author} {\bibfnamefont {S.}~\bibnamefont {Fritzsche}},\ }\href
  {https://doi.org/10.1103/PhysRevA.94.033420} {\bibfield  {journal} {\bibinfo
  {journal} {Phys. Rev. A}\ }\textbf {\bibinfo {volume} {94}},\ \bibinfo
  {pages} {033420} (\bibinfo {year} {2016})}\BibitemShut {NoStop}%
\bibitem [{\citenamefont {Karlovets}\ \emph {et~al.}(2017)\citenamefont
  {Karlovets}, \citenamefont {Kotkin}, \citenamefont {Serbo},\ and\
  \citenamefont {Surzhykov}}]{karlovets2017}%
  \BibitemOpen
  \bibfield  {author} {\bibinfo {author} {\bibfnamefont {D.~V.}\ \bibnamefont
  {Karlovets}}, \bibinfo {author} {\bibfnamefont {G.~L.}\ \bibnamefont
  {Kotkin}}, \bibinfo {author} {\bibfnamefont {V.~G.}\ \bibnamefont {Serbo}},\
  and\ \bibinfo {author} {\bibfnamefont {A.}~\bibnamefont {Surzhykov}},\ }\href
  {https://doi.org/10.1103/PhysRevA.95.032703} {\bibfield  {journal} {\bibinfo
  {journal} {Phys. Rev. A}\ }\textbf {\bibinfo {volume} {95}},\ \bibinfo
  {pages} {032703} (\bibinfo {year} {2017})}\BibitemShut {NoStop}%
\bibitem [{\citenamefont {Babiker}\ \emph {et~al.}(2018)\citenamefont
  {Babiker}, \citenamefont {Andrews},\ and\ \citenamefont
  {Lembessis}}]{LightandAtom}%
  \BibitemOpen
  \bibfield  {author} {\bibinfo {author} {\bibfnamefont {M.}~\bibnamefont
  {Babiker}}, \bibinfo {author} {\bibfnamefont {D.~L.}\ \bibnamefont
  {Andrews}},\ and\ \bibinfo {author} {\bibfnamefont {V.~E.}\ \bibnamefont
  {Lembessis}},\ }\href {https://doi.org/10.1088/2040-8986/aaed14} {\bibfield
  {journal} {\bibinfo  {journal} {Journal of Optics}\ }\textbf {\bibinfo
  {volume} {21}},\ \bibinfo {pages} {013001} (\bibinfo {year}
  {2018})}\BibitemShut {NoStop}%
\bibitem [{\citenamefont {Peshkov}\ \emph {et~al.}(2018)\citenamefont
  {Peshkov}, \citenamefont {Volotka}, \citenamefont {Surzhykov},\ and\
  \citenamefont {Fritzsche}}]{peshkov2018}%
  \BibitemOpen
  \bibfield  {author} {\bibinfo {author} {\bibfnamefont {A.~A.}\ \bibnamefont
  {Peshkov}}, \bibinfo {author} {\bibfnamefont {A.~V.}\ \bibnamefont
  {Volotka}}, \bibinfo {author} {\bibfnamefont {A.}~\bibnamefont {Surzhykov}},\
  and\ \bibinfo {author} {\bibfnamefont {S.}~\bibnamefont {Fritzsche}},\ }\href
  {https://doi.org/10.1103/PhysRevA.97.023802} {\bibfield  {journal} {\bibinfo
  {journal} {Phys. Rev. A}\ }\textbf {\bibinfo {volume} {97}},\ \bibinfo
  {pages} {023802} (\bibinfo {year} {2018})}\BibitemShut {NoStop}%
\bibitem [{\citenamefont {Afanasev}\ \emph {et~al.}(2018)\citenamefont
  {Afanasev}, \citenamefont {Carlson}, \citenamefont {Schmiegelow},
  \citenamefont {Schulz}, \citenamefont {Schmidt-Kaler},\ and\ \citenamefont
  {Solyanik}}]{afanasev2018}%
  \BibitemOpen
  \bibfield  {author} {\bibinfo {author} {\bibfnamefont {A.}~\bibnamefont
  {Afanasev}}, \bibinfo {author} {\bibfnamefont {C.~E.}\ \bibnamefont
  {Carlson}}, \bibinfo {author} {\bibfnamefont {C.~T.}\ \bibnamefont
  {Schmiegelow}}, \bibinfo {author} {\bibfnamefont {J.}~\bibnamefont {Schulz}},
  \bibinfo {author} {\bibfnamefont {F.}~\bibnamefont {Schmidt-Kaler}},\ and\
  \bibinfo {author} {\bibfnamefont {M.}~\bibnamefont {Solyanik}},\ }\href
  {https://doi.org/10.1088/1367-2630/aaa63d} {\bibfield  {journal} {\bibinfo
  {journal} {New J. Phys.}\ }\textbf {\bibinfo {volume} {20}},\ \bibinfo
  {pages} {023032} (\bibinfo {year} {2018})}\BibitemShut {NoStop}%
\bibitem [{\citenamefont {Zaytsev}\ \emph {et~al.}(2018)\citenamefont
  {Zaytsev}, \citenamefont {Surzhykov}, \citenamefont {Shabaev},\ and\
  \citenamefont {St\"ohlker}}]{zaytsev2018}%
  \BibitemOpen
  \bibfield  {author} {\bibinfo {author} {\bibfnamefont {V.~A.}\ \bibnamefont
  {Zaytsev}}, \bibinfo {author} {\bibfnamefont {A.~S.}\ \bibnamefont
  {Surzhykov}}, \bibinfo {author} {\bibfnamefont {V.~M.}\ \bibnamefont
  {Shabaev}},\ and\ \bibinfo {author} {\bibfnamefont {T.}~\bibnamefont
  {St\"ohlker}},\ }\href {https://doi.org/10.1103/PhysRevA.97.043808}
  {\bibfield  {journal} {\bibinfo  {journal} {Phys. Rev. A}\ }\textbf {\bibinfo
  {volume} {97}},\ \bibinfo {pages} {043808} (\bibinfo {year}
  {2018})}\BibitemShut {NoStop}%
\bibitem [{\citenamefont {Forbes}(2019)}]{forbes2019}%
  \BibitemOpen
  \bibfield  {author} {\bibinfo {author} {\bibfnamefont {K.~A.}\ \bibnamefont
  {Forbes}},\ }\href {https://doi.org/10.1103/PhysRevLett.122.103201}
  {\bibfield  {journal} {\bibinfo  {journal} {Phys. Rev. Lett.}\ }\textbf
  {\bibinfo {volume} {122}},\ \bibinfo {pages} {103201} (\bibinfo {year}
  {2019})}\BibitemShut {NoStop}%
\bibitem [{\citenamefont {Afanasev}\ \emph {et~al.}(2020)\citenamefont
  {Afanasev}, \citenamefont {Carlson},\ and\ \citenamefont
  {Wang}}]{afanasev2020}%
  \BibitemOpen
  \bibfield  {author} {\bibinfo {author} {\bibfnamefont {A.}~\bibnamefont
  {Afanasev}}, \bibinfo {author} {\bibfnamefont {C.~E.}\ \bibnamefont
  {Carlson}},\ and\ \bibinfo {author} {\bibfnamefont {H.}~\bibnamefont
  {Wang}},\ }\href {https://doi.org/10.1088/2040-8986/ab8288} {\bibfield
  {journal} {\bibinfo  {journal} {J. Opt.}\ }\textbf {\bibinfo {volume} {22}},\
  \bibinfo {pages} {054001} (\bibinfo {year} {2020})}\BibitemShut {NoStop}%
\bibitem [{\citenamefont {Schulz}\ \emph {et~al.}(2020)\citenamefont {Schulz},
  \citenamefont {Peshkov}, \citenamefont {M\"uller}, \citenamefont {Lange},
  \citenamefont {Huntemann}, \citenamefont {Tamm}, \citenamefont {Peik},\ and\
  \citenamefont {Surzhykov}}]{schulz2020}%
  \BibitemOpen
  \bibfield  {author} {\bibinfo {author} {\bibfnamefont {S.~A.-L.}\
  \bibnamefont {Schulz}}, \bibinfo {author} {\bibfnamefont {A.~A.}\
  \bibnamefont {Peshkov}}, \bibinfo {author} {\bibfnamefont {R.~A.}\
  \bibnamefont {M\"uller}}, \bibinfo {author} {\bibfnamefont {R.}~\bibnamefont
  {Lange}}, \bibinfo {author} {\bibfnamefont {N.}~\bibnamefont {Huntemann}},
  \bibinfo {author} {\bibfnamefont {C.}~\bibnamefont {Tamm}}, \bibinfo {author}
  {\bibfnamefont {E.}~\bibnamefont {Peik}},\ and\ \bibinfo {author}
  {\bibfnamefont {A.}~\bibnamefont {Surzhykov}},\ }\href
  {https://doi.org/10.1103/PhysRevA.102.012812} {\bibfield  {journal} {\bibinfo
   {journal} {Phys. Rev. A}\ }\textbf {\bibinfo {volume} {102}},\ \bibinfo
  {pages} {012812} (\bibinfo {year} {2020})}\BibitemShut {NoStop}%
\bibitem [{\citenamefont {Ivanov}(2022)}]{ivanov2022}%
  \BibitemOpen
  \bibfield  {author} {\bibinfo {author} {\bibfnamefont {I.~P.}\ \bibnamefont
  {Ivanov}},\ }\href {https://doi.org/https://doi.org/10.1002/andp.202100128}
  {\bibfield  {journal} {\bibinfo  {journal} {Annalen der Physik}\ }\textbf
  {\bibinfo {volume} {534}},\ \bibinfo {pages} {2100128} (\bibinfo {year}
  {2022})}\BibitemShut {NoStop}%
\bibitem [{\citenamefont {Serbo}\ \emph {et~al.}(2022)\citenamefont {Serbo},
  \citenamefont {Surzhykov},\ and\ \citenamefont {Volotka}}]{serbo2022}%
  \BibitemOpen
  \bibfield  {author} {\bibinfo {author} {\bibfnamefont {V.~G.}\ \bibnamefont
  {Serbo}}, \bibinfo {author} {\bibfnamefont {A.}~\bibnamefont {Surzhykov}},\
  and\ \bibinfo {author} {\bibfnamefont {A.}~\bibnamefont {Volotka}},\ }\href
  {https://doi.org/https://doi.org/10.1002/andp.202100199} {\bibfield
  {journal} {\bibinfo  {journal} {Annalen der Physik}\ }\textbf {\bibinfo
  {volume} {534}},\ \bibinfo {pages} {2100199} (\bibinfo {year}
  {2022})}\BibitemShut {NoStop}%
\bibitem [{\citenamefont {Schmiegelow}\ \emph {et~al.}(2016)\citenamefont
  {Schmiegelow}, \citenamefont {Schulz}, \citenamefont {Kaufmann},
  \citenamefont {Ruster}, \citenamefont {Poschinger},\ and\ \citenamefont
  {Schmidt-Kaler}}]{schmiegelow2016}%
  \BibitemOpen
  \bibfield  {author} {\bibinfo {author} {\bibfnamefont {C.}~\bibnamefont
  {Schmiegelow}}, \bibinfo {author} {\bibfnamefont {J.}~\bibnamefont {Schulz}},
  \bibinfo {author} {\bibfnamefont {H.}~\bibnamefont {Kaufmann}}, \bibinfo
  {author} {\bibfnamefont {T.}~\bibnamefont {Ruster}}, \bibinfo {author}
  {\bibfnamefont {U.~G.}\ \bibnamefont {Poschinger}},\ and\ \bibinfo {author}
  {\bibfnamefont {F.}~\bibnamefont {Schmidt-Kaler}},\ }\href
  {https://doi.org/doi.org/10.1038/ncomms12998} {\bibfield  {journal} {\bibinfo
   {journal} {Nat. Commun.}\ }\textbf {\bibinfo {volume} {7}},\ \bibinfo
  {pages} {12998} (\bibinfo {year} {2016})}\BibitemShut {NoStop}%
\bibitem [{\citenamefont {Solyanik-Gorgone}\ \emph {et~al.}(2019)\citenamefont
  {Solyanik-Gorgone}, \citenamefont {Afanasev}, \citenamefont {Carlson},
  \citenamefont {Schmiegelow},\ and\ \citenamefont
  {Schmidt-Kaler}}]{ForbidTrans1}%
  \BibitemOpen
  \bibfield  {author} {\bibinfo {author} {\bibfnamefont {M.}~\bibnamefont
  {Solyanik-Gorgone}}, \bibinfo {author} {\bibfnamefont {A.}~\bibnamefont
  {Afanasev}}, \bibinfo {author} {\bibfnamefont {C.~E.}\ \bibnamefont
  {Carlson}}, \bibinfo {author} {\bibfnamefont {C.~T.}\ \bibnamefont
  {Schmiegelow}},\ and\ \bibinfo {author} {\bibfnamefont {F.}~\bibnamefont
  {Schmidt-Kaler}},\ }\href {https://doi.org/10.1364/JOSAB.36.000565}
  {\bibfield  {journal} {\bibinfo  {journal} {J. Opt. Soc. Am. B}\ }\textbf
  {\bibinfo {volume} {36}},\ \bibinfo {pages} {565} (\bibinfo {year}
  {2019})}\BibitemShut {NoStop}%
\bibitem [{\citenamefont {Davis}\ \emph {et~al.}(2013)\citenamefont {Davis},
  \citenamefont {Kaplan},\ and\ \citenamefont {McGuire}}]{Davis_2013}%
  \BibitemOpen
  \bibfield  {author} {\bibinfo {author} {\bibfnamefont {B.~S.}\ \bibnamefont
  {Davis}}, \bibinfo {author} {\bibfnamefont {L.}~\bibnamefont {Kaplan}},\ and\
  \bibinfo {author} {\bibfnamefont {J.~H.}\ \bibnamefont {McGuire}},\ }\href
  {https://doi.org/10.1088/2040-8978/15/3/035403} {\bibfield  {journal}
  {\bibinfo  {journal} {Journal of Optics}\ }\textbf {\bibinfo {volume} {15}},\
  \bibinfo {pages} {035403} (\bibinfo {year} {2013})}\BibitemShut {NoStop}%
\bibitem [{\citenamefont {Stopp}\ \emph {et~al.}(2022)\citenamefont {Stopp},
  \citenamefont {Verde}, \citenamefont {Katz}, \citenamefont {Drechsler},
  \citenamefont {Schmiegelow},\ and\ \citenamefont
  {Schmidt-Kaler}}]{stopp2022}%
  \BibitemOpen
  \bibfield  {author} {\bibinfo {author} {\bibfnamefont {F.}~\bibnamefont
  {Stopp}}, \bibinfo {author} {\bibfnamefont {M.}~\bibnamefont {Verde}},
  \bibinfo {author} {\bibfnamefont {M.}~\bibnamefont {Katz}}, \bibinfo {author}
  {\bibfnamefont {M.}~\bibnamefont {Drechsler}}, \bibinfo {author}
  {\bibfnamefont {C.~T.}\ \bibnamefont {Schmiegelow}},\ and\ \bibinfo {author}
  {\bibfnamefont {F.}~\bibnamefont {Schmidt-Kaler}},\ }\href
  {https://doi.org/10.1103/PhysRevLett.129.263603} {\bibfield  {journal}
  {\bibinfo  {journal} {Phys. Rev. Lett.}\ }\textbf {\bibinfo {volume} {129}},\
  \bibinfo {pages} {263603} (\bibinfo {year} {2022})}\BibitemShut {NoStop}%
\bibitem [{\citenamefont {Peshkov}\ \emph {et~al.}(2023)\citenamefont
  {Peshkov}, \citenamefont {Bidasyuk}, \citenamefont {Lange}, \citenamefont
  {Huntemann}, \citenamefont {Peik},\ and\ \citenamefont
  {Surzhykov}}]{peshkov2023}%
  \BibitemOpen
  \bibfield  {author} {\bibinfo {author} {\bibfnamefont {A.~A.}\ \bibnamefont
  {Peshkov}}, \bibinfo {author} {\bibfnamefont {Y.~M.}\ \bibnamefont
  {Bidasyuk}}, \bibinfo {author} {\bibfnamefont {R.}~\bibnamefont {Lange}},
  \bibinfo {author} {\bibfnamefont {N.}~\bibnamefont {Huntemann}}, \bibinfo
  {author} {\bibfnamefont {E.}~\bibnamefont {Peik}},\ and\ \bibinfo {author}
  {\bibfnamefont {A.}~\bibnamefont {Surzhykov}},\ }\href
  {https://doi.org/10.1103/PhysRevA.107.023106} {\bibfield  {journal} {\bibinfo
   {journal} {Phys. Rev. A}\ }\textbf {\bibinfo {volume} {107}},\ \bibinfo
  {pages} {023106} (\bibinfo {year} {2023})}\BibitemShut {NoStop}%
\bibitem [{\citenamefont {Luski}\ \emph {et~al.}(2021)\citenamefont {Luski},
  \citenamefont {Segev}, \citenamefont {David}, \citenamefont {Bitton},
  \citenamefont {Nadler}, \citenamefont {Barnea}, \citenamefont {Gorlach},
  \citenamefont {Cheshnovsky}, \citenamefont {Kaminer},\ and\ \citenamefont
  {Narevicius}}]{VortAt2021}%
  \BibitemOpen
  \bibfield  {author} {\bibinfo {author} {\bibfnamefont {A.}~\bibnamefont
  {Luski}}, \bibinfo {author} {\bibfnamefont {Y.}~\bibnamefont {Segev}},
  \bibinfo {author} {\bibfnamefont {R.}~\bibnamefont {David}}, \bibinfo
  {author} {\bibfnamefont {O.}~\bibnamefont {Bitton}}, \bibinfo {author}
  {\bibfnamefont {H.}~\bibnamefont {Nadler}}, \bibinfo {author} {\bibfnamefont
  {A.~R.}\ \bibnamefont {Barnea}}, \bibinfo {author} {\bibfnamefont
  {A.}~\bibnamefont {Gorlach}}, \bibinfo {author} {\bibfnamefont
  {O.}~\bibnamefont {Cheshnovsky}}, \bibinfo {author} {\bibfnamefont
  {I.}~\bibnamefont {Kaminer}},\ and\ \bibinfo {author} {\bibfnamefont
  {E.}~\bibnamefont {Narevicius}},\ }\href
  {https://doi.org/10.1126/science.abj2451} {\bibfield  {journal} {\bibinfo
  {journal} {Science}\ }\textbf {\bibinfo {volume} {373}},\ \bibinfo {pages}
  {1105} (\bibinfo {year} {2021})}\BibitemShut {NoStop}%
\bibitem [{\citenamefont {Bethe}\ and\ \citenamefont {Salpeter}(1957)}]{bethe}%
  \BibitemOpen
  \bibfield  {author} {\bibinfo {author} {\bibfnamefont {H.~A.}\ \bibnamefont
  {Bethe}}\ and\ \bibinfo {author} {\bibfnamefont {E.~E.}\ \bibnamefont
  {Salpeter}},\ }\href@noop {} {\emph {\bibinfo {title} {Quantum Mechanics of
  One- and Two-Electron Atoms}}}\ (\bibinfo  {publisher} {{Springer-Verlag}},\
  \bibinfo {year} {1957})\BibitemShut {NoStop}%
\bibitem [{\citenamefont {Berestetskii}\ \emph {et~al.}(1982)\citenamefont
  {Berestetskii}, \citenamefont {Lifshitz},\ and\ \citenamefont
  {Pitaevskii}}]{BLP}%
  \BibitemOpen
  \bibfield  {author} {\bibinfo {author} {\bibfnamefont {V.}~\bibnamefont
  {Berestetskii}}, \bibinfo {author} {\bibfnamefont {E.}~\bibnamefont
  {Lifshitz}},\ and\ \bibinfo {author} {\bibfnamefont {L.}~\bibnamefont
  {Pitaevskii}},\ }\href@noop {} {\emph {\bibinfo {title} {Quantum
  Electrodynamics}}}\ (\bibinfo  {publisher} {Oxford: Pergamon},\ \bibinfo
  {year} {1982})\BibitemShut {NoStop}%
\bibitem [{\citenamefont {Shabaev}(2001)}]{Shabaev2001}%
  \BibitemOpen
  \bibfield  {author} {\bibinfo {author} {\bibfnamefont {V.~M.}\ \bibnamefont
  {Shabaev}},\ }\href {https://doi.org/10.1103/PhysRevA.64.052104} {\bibfield
  {journal} {\bibinfo  {journal} {Phys. Rev. A}\ }\textbf {\bibinfo {volume}
  {64}},\ \bibinfo {pages} {052104} (\bibinfo {year} {2001})}\BibitemShut
  {NoStop}%
\bibitem [{\citenamefont {Fried}\ and\ \citenamefont
  {Martin}(1963)}]{fried:1963:574}%
  \BibitemOpen
  \bibfield  {author} {\bibinfo {author} {\bibfnamefont {Z.}~\bibnamefont
  {Fried}}\ and\ \bibinfo {author} {\bibfnamefont {A.~D.}\ \bibnamefont
  {Martin}},\ }\href@noop {} {\bibfield  {journal} {\bibinfo  {journal} {Nuovo
  Cimento}\ }\textbf {\bibinfo {volume} {29}},\ \bibinfo {pages} {574}
  (\bibinfo {year} {1963})}\BibitemShut {NoStop}%
\bibitem [{\citenamefont {Bacher}(1984)}]{bacher:1984:135}%
  \BibitemOpen
  \bibfield  {author} {\bibinfo {author} {\bibfnamefont {R.}~\bibnamefont
  {Bacher}},\ }\href@noop {} {\bibfield  {journal} {\bibinfo  {journal} {Z.
  Phys. A}\ }\textbf {\bibinfo {volume} {315}},\ \bibinfo {pages} {135}
  (\bibinfo {year} {1984})}\BibitemShut {NoStop}%
\bibitem [{\citenamefont {Karshenboim}(1997)}]{karshenboim:1997:4311}%
  \BibitemOpen
  \bibfield  {author} {\bibinfo {author} {\bibfnamefont {S.~G.}\ \bibnamefont
  {Karshenboim}},\ }\href@noop {} {\bibfield  {journal} {\bibinfo  {journal}
  {Phys. Rev. A}\ }\textbf {\bibinfo {volume} {56}},\ \bibinfo {pages} {4311}
  (\bibinfo {year} {1997})}\BibitemShut {NoStop}%
\bibitem [{\citenamefont {Shabaev}(1998)}]{shabaev:1998:907}%
  \BibitemOpen
  \bibfield  {author} {\bibinfo {author} {\bibfnamefont {V.~M.}\ \bibnamefont
  {Shabaev}},\ }\href@noop {} {\bibfield  {journal} {\bibinfo  {journal} {Can.
  J. Phys.}\ }\textbf {\bibinfo {volume} {76}},\ \bibinfo {pages} {907}
  (\bibinfo {year} {1998})}\BibitemShut {NoStop}%
\bibitem [{\citenamefont {Pachucki}(2003)}]{pachucki:2003:012504}%
  \BibitemOpen
  \bibfield  {author} {\bibinfo {author} {\bibfnamefont {K.}~\bibnamefont
  {Pachucki}},\ }\href@noop {} {\bibfield  {journal} {\bibinfo  {journal}
  {Phys. Rev. A}\ }\textbf {\bibinfo {volume} {67}},\ \bibinfo {pages} {012504}
  (\bibinfo {year} {2003})}\BibitemShut {NoStop}%
\bibitem [{\citenamefont {Volotka}\ \emph {et~al.}(2008)\citenamefont
  {Volotka}, \citenamefont {Glazov}, \citenamefont {Plunien}, \citenamefont
  {Shabaev},\ and\ \citenamefont {Tupitsyn}}]{volotka:2008:167}%
  \BibitemOpen
  \bibfield  {author} {\bibinfo {author} {\bibfnamefont {A.~V.}\ \bibnamefont
  {Volotka}}, \bibinfo {author} {\bibfnamefont {D.~A.}\ \bibnamefont {Glazov}},
  \bibinfo {author} {\bibfnamefont {G.}~\bibnamefont {Plunien}}, \bibinfo
  {author} {\bibfnamefont {V.~M.}\ \bibnamefont {Shabaev}},\ and\ \bibinfo
  {author} {\bibfnamefont {I.~I.}\ \bibnamefont {Tupitsyn}},\ }\href@noop {}
  {\bibfield  {journal} {\bibinfo  {journal} {Eur. Phys. J. D}\ }\textbf
  {\bibinfo {volume} {48}},\ \bibinfo {pages} {167} (\bibinfo {year}
  {2008})}\BibitemShut {NoStop}%
\bibitem [{\citenamefont {Bondy}\ \emph {et~al.}(2020)\citenamefont {Bondy},
  \citenamefont {Morton},\ and\ \citenamefont {Drake}}]{bondy:2020:052807}%
  \BibitemOpen
  \bibfield  {author} {\bibinfo {author} {\bibfnamefont {A.~T.}\ \bibnamefont
  {Bondy}}, \bibinfo {author} {\bibfnamefont {D.~C.}\ \bibnamefont {Morton}},\
  and\ \bibinfo {author} {\bibfnamefont {G.~W.~F.}\ \bibnamefont {Drake}},\
  }\href {https://doi.org/10.1103/PhysRevA.102.052807} {\bibfield  {journal}
  {\bibinfo  {journal} {Phys. Rev. A}\ }\textbf {\bibinfo {volume} {102}},\
  \bibinfo {pages} {052807} (\bibinfo {year} {2020})}\BibitemShut {NoStop}%
\bibitem [{\citenamefont {Pachucki}(2007)}]{pachucki:2008}%
  \BibitemOpen
  \bibfield  {author} {\bibinfo {author} {\bibfnamefont {K.}~\bibnamefont
  {Pachucki}},\ }\href {https://doi.org/10.1103/PhysRevA.76.022106} {\bibfield
  {journal} {\bibinfo  {journal} {Phys. Rev. A}\ }\textbf {\bibinfo {volume}
  {76}},\ \bibinfo {pages} {022106} (\bibinfo {year} {2007})}\BibitemShut
  {NoStop}%
\bibitem [{\citenamefont {Peskin}\ and\ \citenamefont
  {Schroeder}(1995)}]{Peskin}%
  \BibitemOpen
  \bibfield  {author} {\bibinfo {author} {\bibfnamefont {M.~E.}\ \bibnamefont
  {Peskin}}\ and\ \bibinfo {author} {\bibfnamefont {D.~V.}\ \bibnamefont
  {Schroeder}},\ }\href@noop {} {\emph {\bibinfo {title} {{An Introduction to
  quantum field theory}}}}\ (\bibinfo  {publisher} {Addison-Wesley},\ \bibinfo
  {address} {Reading, USA},\ \bibinfo {year} {1995})\BibitemShut {NoStop}%
\bibitem [{\citenamefont {Jentschura}\ and\ \citenamefont
  {Serbo}(2011)}]{Jentschura:2011aa}%
  \BibitemOpen
  \bibfield  {author} {\bibinfo {author} {\bibfnamefont {U.~D.}\ \bibnamefont
  {Jentschura}}\ and\ \bibinfo {author} {\bibfnamefont {V.~G.}\ \bibnamefont
  {Serbo}},\ }\href {https://doi.org/10.1140/epjc/s10052-011-1571-z} {\bibfield
   {journal} {\bibinfo  {journal} {The European Physical Journal C}\ }\textbf
  {\bibinfo {volume} {71}},\ \bibinfo {pages} {1571} (\bibinfo {year}
  {2011})}\BibitemShut {NoStop}%
\bibitem [{\citenamefont {Ivanov}(2011)}]{Ivanov2011}%
  \BibitemOpen
  \bibfield  {author} {\bibinfo {author} {\bibfnamefont {I.~P.}\ \bibnamefont
  {Ivanov}},\ }\href {https://doi.org/10.1103/PhysRevD.83.093001} {\bibfield
  {journal} {\bibinfo  {journal} {Phys. Rev. D}\ }\textbf {\bibinfo {volume}
  {83}},\ \bibinfo {pages} {093001} (\bibinfo {year} {2011})}\BibitemShut
  {NoStop}%
\end{thebibliography}%
\end{document}